\documentclass[preprint]{aastex}
\tighten

\input psfig

\def\kms{\,{\rm km\, s^{-1}} }
\def\kmsm{\,{\rm km\, s^{-1}\, Mpc^{-1}} }

\def\kpc{\,{\rm kpc} }

\begin{document}

\title{The Evolution of a Mass-Selected Sample of Early-Type Field
Galaxies} 

\author{D. Rusin\altaffilmark{1}, C.S. Kochanek\altaffilmark{1}}
\author{E.E. Falco\altaffilmark{2}, C.R. Keeton\altaffilmark{3},
B.A. McLeod\altaffilmark{1}}
\author{C.D. Impey\altaffilmark{4}, J. Leh\'ar\altaffilmark{1},
J.A. Mu\~noz\altaffilmark{5}, C.Y. Peng\altaffilmark{4},
H.-W. Rix\altaffilmark{6}}

\altaffiltext{1}{Harvard-Smithsonian Center for Astrophysics, 60
Garden St., Cambridge, MA 02138}

\altaffiltext{2}{Smithsonian Institution, F.L. Whipple Observatory,
P.O. Box 97, 670 Mount Hopkins, Amado, AZ 85645}

\altaffiltext{3}{Astronomy and Astrophysics Department, University of
Chicago, 5640 S. Ellis Ave., Chicago, IL 60637}

\altaffiltext{4}{Steward Observatory, University of Arizona, Tucson,
AZ 85721}

\altaffiltext{5}{Instituto de Astrofisica de Canarias, Via Lactea,
E38200 La Laguna, Tenerife, Spain}

\altaffiltext{6}{Max-Planck-Institut fuer Astronomie, Koenigsstuhl 17,
D-69117 Heidelberg, Germany}

\begin{abstract}

We investigate the evolution of mass-selected early-type field
galaxies using a sample of 28 gravitational lenses spanning the
redshift range $0 \la z \la 1$. Based on the redshift-dependent
intercept of the fundamental plane in the rest frame $B$ band, we
measure an evolution rate of $d\log (M/L)_B / dz = -0.56 \pm 0.04$
(all errors are $1 \sigma$ unless noted) if we directly compare to the
local intercept measured from the Coma cluster. Re-fitting the local
intercept helps minimize potential systematic errors, and yields an
evolution rate of $d\log (M/L)_B / dz = -0.54 \pm 0.09$. An evolution
analysis of properly-corrected aperture mass-to-light ratios (defined
by the lensed image separations) is closely related to the
Faber-Jackson relation. In rest frame $B$ band we find an evolution
rate of $d\log (M/L)_B / dz = -0.41 \pm 0.21$, a present-day
characteristic magnitude of $M_{*0} = -19.70 + 5 \log h \pm 0.29$
(assuming a characteristic velocity dispersion of $\sigma_{DM*} = 225
\kms$), and a Faber-Jackson slope of $\gamma_{FJ} = 3.29 \pm
0.58$. The measured evolution rates favor old stellar populations
(mean formation redshift $\langle z_f \rangle > 1.8$ at $2\sigma$
confidence for a Salpeter initial mass function and a flat $\Omega_m
=0.3$ cosmology) among early-type field galaxies, and argue against
significant episodes of star formation at $z < 1$.

\end{abstract}

\keywords{galaxies: elliptical and lenticular, cD -- galaxies: evolution
-- gravitational lensing}

\section{Introduction} 

In recent years, enormous progress has been made in tracing the
evolution of the formation rate of massive stars with cosmic epoch
(e.g., Madau, Pozzetti \& Dickinson 1998 and references therein).
Although these massive stars initially dominate the luminosity of
stellar systems, particularly the ionizing luminosity, they represent
a negligible fraction of the overall stellar mass in galaxies. Tracing
the formation history of the lower mass stars is more difficult, but
it is necessary if we are to have a complete picture of the
star-formation history of the universe.  One means for tracing the
formation history of these stars is to follow the evolution of the
mass-to-light ratios of galaxies with redshift. Because studying
evolution using mass-to-light ratios requires a determination of the
galaxy mass or a surrogate for the mass, this approach has developed
slowly due to the difficulty in measuring the rotation curves (e.g.,
Vogt et al.\ 1996; Ziegler et al.\ 2002) or velocity dispersions
(e.g., van Dokkum \& Franx 1996) of galaxies at intermediate redshift.

Most of the progress has been made in studies of early-type galaxies.
It has long been known that the remarkable homogeneity of these
galaxies is a powerful constraint on formation models.  First, the
population exhibits very uniform colors both locally (e.g., Bower,
Lucey \& Ellis 1992) and at $z \sim 1$ (e.g., Ellis et al.\ 1997;
Stanford, Eisenhardt \& Dickinson 1998).  Second, the early-type
galaxies follow a tight correlation among central velocity dispersion,
effective (half-light) radius, and surface brightness known as the
fundamental plane (FP; Djorgovski \& Davis 1987; Dressler et al.\
1987). The scatter in the FP, which is closely related to the scatter
in mass-to-light ratio, is locally small (e.g., Jorgensen, Franx \&
Kjaergaard 1996; Pahre, Djorgovski \& de Carvalho 1998a; Bernardi et
al.\ 2001), does not evolve significantly with redshift (e.g., van
Dokkum \& Franx 1996; Kelson et al.\ 1997; Pahre, Djorgovski \& de
Carvalho 2001), and shows little dependence on the environment (e.g.,
van Dokkum et al.\ 1998, 2001; Kochanek et al.\ 2000; Treu et al.\
2001).

In modern theoretical models of galaxy formation, (e.g., White \& Rees
1978; Davis et al.\ 1985; Kauffmann, White \& Guiderdoni 1993; Cole et
al.\ 1994; Kauffmann 1996), early-type galaxies are assembled via the
mergers of late-type galaxies. Support for the hierarchical clustering
model is provided by the observation that high-redshift clusters
exhibit both increased merger rates (e.g., Lavery \& Henry 1988;
Dressler et al.\ 1994; Couch et al.\ 1998; van Dokkum et al.\ 1999,
2000) and decreased fractions of elliptical galaxies (e.g., Dressler
et al.\ 1997; Couch et al.\ 1998; van Dokkum et al.\ 2000; see
Fabricant, Franx \& van Dokkum 2000 for insightful criticism) compared
to their present-day counterparts.  Semi-analytic CDM models (Baugh,
Cole \& Frenk 1996; Kauffmann 1996; Kauffman \& Charlot 1998) further
predict that early-type galaxies in the field should contain
recently-formed ($z_f<1$) stellar populations, while those in clusters
should have significantly older stellar populations. Local
observational studies have difficulty separating the effects of age
and metallicity (e.g., Faber 1973; Worthey 1994; Trager et al.\ 2000),
but such degeneracies can be broken by measuring the evolution of
mass-to-light ratios with redshift.

Morphological evolution can complicate attempts to study the stellar
evolution if the number density of massive early-type galaxies evolves
rapidly to redshift unity (e.g., Kauffmann 1996; van Dokkum \& Franx
2001). For example, a paucity of E/S0 galaxies with young blue stars
out to $z\sim 1$ may be produced by several different star-formation
and merger histories. First, all early-type galaxies could have been
assembled at very high redshift (e.g., $z\sim 3$). Here the stars of
$z<1$ E/S0 galaxies look old because the galaxies themselves are old.
Second, some early-type galaxies could have been assembled more
recently (e.g., $z < 1$) from galaxies that had old stars and no
ongoing star-formation prior to the merger (e.g., van Dokkum et al.\
1999, 2000). The assembly age of the E/S0 galaxies would be much lower
than the stellar age in this scenario, but the stellar age derived
from early-types out to $z\sim 1$ would accurately represent the
stellar age among all progenitors of present-day early-type
galaxies. Finally, some E/S0s could be assembled at $z<1$ through the
mergers of star-forming late-type galaxies. In this case, a range of
assembly times leads to an evolving distribution of stellar formation
times for the early-type population.  As one moves to higher
redshifts, the younger present-day early-type galaxies have yet to be
assembled from their smaller, star-forming progenitor galaxies and do
not appear in the higher redshift E/S0 samples. Consequently, by
considering only early-type galaxies out to $z\sim 1$, the sample will
be biased toward the oldest progenitors of the present-day population,
leading to an under-estimate of the luminosity evolution rate and an
over-estimate of the mean stellar age. The possible magnitude of the
progenitor bias in clusters is investigated in detail by van Dokkum \&
Franx (2001), who demonstrate that the observed evolution may be
under-estimated by as much as $\Delta \log (M/L)_B \sim 0.1z$ in
cluster environments. The true effect of progenitor bias may be much
smaller, however, if E/S0s are being formed from the mergers of old
galaxies without ongoing star-formation, a scenario suggested by the
``red mergers'' observed in MS1054--03 (van Dokkum et al.\ 1999). In
addition, it would not be unreasonable to suggest that progenitor bias
effects may be smaller in the field than in high-density environments
such as clusters. Recent evidence against a strong evolution in the
number density of luminous early-type field galaxies out to $z\sim 1$
(e.g., Schade et al.\ 1999; Im et al.\ 2002) supports this scenario,
and implies that high-redshift early-type galaxies are indeed the
progenitors of the local population.  Clearly both photometric and
morphological studies are necessary to obtain a complete picture of
early-type galaxy evolution. For now, we will focus on the former.

Empirical relations such as the FP (Djorgovski \& Davis 1987; Dressler
et al.\ 1987; \S 3) and to a lesser extent, the Faber-Jackson rule
(FJ; 1976; \S 4), are crucial to the investigation of luminosity
evolution. Each allows for the derivation of a mass-related scale by
which to normalize the photometric data, thereby eliminating the
ambiguities in evolution studies based on luminosity functions (e.g.,
Lilly et al.\ 1995; Lin et al.\ 1999; Im et al.\ 2002). In general,
the FP is preferred because its scatter is much smaller than that of
the FJ (e.g., Pahre et al.\ 1998a). The evolution of both cluster
(e.g., van Dokkum et al.\ 1998; Jorgensen et al.\ 1999; Pahre et al.\
2001) and field samples (e.g., Kochanek et al.\ 2000; van Dokkum et
al.\ 2001; Treu et al.\ 2001, 2002) have been investigated within the
framework of the fundamental plane. Recent results in the rest frame
$B$ band are as follows. Compiling data from five clusters (Jorgensen
et al.\ 1996; van Dokkum \& Franx 1996; Kelson et al.\ 1997; van
Dokkum et al.\ 1998) spanning the redshift range of $0.02 < z < 0.83$,
van Dokkum et al.\ (1998) measure an evolution rate of $d \log (M/L)_B
/ dz = -0.49 \pm 0.05$ ($\Omega_m = 0.3, \Omega_{\Lambda} = 0.7$) for
cluster early-types. Using 18 early-type field galaxies at
intermediate redshift ($0.15 < z < 0.55$), van Dokkum et al.\ (2001)
find $d \log (M/L)_B / dz = -0.59 \pm 0.15$. These results favor older
stellar populations for both field ($\langle z_f \rangle > 1.5$) and
cluster ($\langle z_f \rangle > 2$) E/S0s, and argue against the large
systematic age difference predicted by semi-analytic galaxy formation
models (e.g., Kauffmann 1996). Alternate conclusions are drawn by Treu
et al.\ (2001, 2002), who find $d \log (M/L)_B / dz =
-0.72^{+0.11}_{-0.16}$ using a sample of 30 early-type field galaxies
in the redshift range $0.10 < z < 0.66$. They interpret the faster
evolution rate as evidence for younger stellar populations among field
ellipticals, perhaps driven by secondary star formation at $z<1$. A
detection of significant [OII] emission in 22\% of their galaxies is
cited in support of these claims.

The greatest potential drawback of the samples of van Dokkum et al.\
(2001) and Treu et al.\ (2001, 2002) is that each is selected on the
basis of luminosity-related properties such as total magnitude,
surface brightness or color. At higher redshifts, all such selection
functions favor intrinsically brighter galaxies, and hence, those with
younger stellar populations and faster evolution rates. Treu et al.\
(2001) discuss the importance of selection effects, and develop a
Bayesian approach to correct their raw results for Malmquist bias.
Even more useful for evolution studies, however, would be a sample of
galaxies selected independently of luminosity-related properties.
Such a sample is provided by gravitational lensing.

Gravitational lenses are the only galaxies selected on the basis of
mass rather than light.  Virtually all lenses are early-type galaxies,
as their large velocity dispersions mean that they provide the
dominant contribution to the lensing optical depth (e.g., Kochanek
1996). Moreover, most lens galaxies reside in low-density environments
-- either the field, or poor groups (e.g., Keeton, Christlein, \&
Zabludoff 2000). While multi-color imaging surveys must overcome
strong surface brightness dimming and Malmquist bias effects to extend
field galaxy samples beyond $z\sim 0.5$, lensing naturally selects a
sample of early-type field galaxies out to $z\sim 1$. Lenses offer a
number of additional advantages for evolution studies. First, because
the lensing mass distributions of early-type galaxies have been
consistently shown to be very close to isothermal (Kochanek 1995; Cohn
et al.\ 2001; Mu\~noz, Kochanek \& Keeton 2001; Rusin et al.\ 2002;
Treu \& Koopmans 2002a; Koopmans \& Treu 2002b), the image separation
provides an accurate estimate of the velocity dispersion. Second, the
aperture defined by the lensed images determines the enclosed
projected mass independently of the mass profile (e.g., Schneider,
Ehlers \& Falco 1992). Third, since the lensing cross-section is
roughly proportional to the mass, the evolution of lens galaxies
traces the {\it mass-averaged} evolution of all early-type galaxies.

Much of the framework for using gravitational lenses to statistically
probe the optical properties of galaxies was laid down by Keeton,
Kochanek \& Falco (1998), particularly with regard to the FJ relation
and aperture mass-to-light ratios. However, their sample was too small
to place robust constraints on evolution.  Kochanek et al.\ (2000)
used improved photometric data to show that lens galaxies can be
matched to the present-day fundamental plane for passively evolving
stellar populations with a mean formation redshift of $\langle z_f
\rangle \ga2$, without explicitly constraining $d\log (M/L) /
dz$. Recent efforts have turned to individual lens systems. The Lens
Structure and Dynamics (LSD) Survey (Koopmans \& Treu 2002a, 2002b;
Treu \& Koopmans 2002a; hereafter collectively ``KT'') has been
measuring the velocity dispersions of gravitational lens galaxies,
with the aim of providing additional constraints on the distribution
of dark matter. The velocity dispersions can be combined with
photometric data from the Hubble Space Telescope (HST) to constrain
the evolution rate. KT have thus far published results for two
galaxies, finding $d \log (M/L)_B/dz = -0.62 \pm 0.08$ for MG2016+112
(Lawrence et al.\ 1984; $z=1.00$) and $d \log (M/L)_B/dz = -0.76 \pm
0.12$ for 0047--281 (Warren et al.\ 1996; $z=0.49$). In each case the
velocity dispersion is consistent with an isothermal mass
distribution.

Ideally one would like to spectroscopically measure the central
velocity dispersions $\sigma_c$ for all lens galaxies, but this will
likely be accomplished in only a limited number of cases. However, the
nearly isothermal mass distribution in early-type lens galaxies allows
us to accurately estimate $\sigma_c$ from image separations (e.g.,
Kochanek 1994; Kochanek et al.\ 2000). Direct measurements of
$\sigma_c$ in several lens systems (e.g., Foltz et al.\ 1992; Leh\'ar
et al.\ 1996; Koopmans \& Treu 2002a, 2002b) have demonstrated that,
on average, the velocity dispersions are well predicted by the
isothermal model. Naturally there are limitations for individual
lenses: if a galaxy has a mass profile that is steeper than
isothermal, then the isothermal assumption will under-estimate the
value of $\sigma_c$ for that particular galaxy, and vice versa. The
gravitational lens PG1115+080 may be one such exception, as its high
stellar velocity dispersion suggests a profile significantly steeper
than those of other lens galaxies (Tonry 1998; Treu \& Koopmans
2002b). But so long as we understand the mean mass profile of lensing
galaxies, our estimation technique should be adequate for
investigating the lens sample statistically, as scatter in the mass
profiles will only lead to scatter in the evolutionary
trends. Therefore, image separations and robust photometry are all we
need to construct a large sample of lens galaxies suitable for
evolution studies.

This paper investigates the evolution of mass-selected early-type
field galaxies using a sample of 28 gravitational lenses spanning the
redshift range $0 \la z \la 1$. Section 2 describes the sample and
outlines our primary data analysis. Section 3 derives the evolution
rate from the fundamental plane. Section 4 investigates evolution
using corrected aperture mass-to-light ratios, a technique that
naturally reduces to a Faber-Jackson formalism.  Section 5 places
constraints on the mean star-formation redshift of mass-selected
early-type galaxies. Section 6 summarizes our findings and discusses
their implications. An appendix investigates the Faber-Jackson
relation in observed bands, and provides a database for estimating the
magnitudes of lens galaxies. We assume a flat $\Omega_m = 0.3$
cosmology, $H_0 = 100 h \kmsm$, and $h = 0.65$ throughout this paper.

\section{Inputs and Assumptions}

\subsection{The Lens Sample}

Gravitational lensing naturally selects a galaxy sample dominated by
massive E/S0 galaxies, as the lensing cross-section scales as the
fourth power of the velocity dispersion (e.g., Turner, Ostriker \&
Gott 1984). This claim has strong observational support, as most lens
galaxies have colors (e.g., Keeton et al.\ 1998; Kochanek et al.\
2000) and spectra (e.g., Fassnacht \& Cohen 1998; Tonry \& Kochanek
1999, 2000; Lubin et al.\ 2000) that are consistent with early-type
morphologies. In fact, only $5$ of the $\sim70$ known arcsecond-scale
gravitational lenses have been shown to be spirals based on visual
morphology, color, spectra or absorption properties. Moreover, when de
Vaucouleurs profiles are fit to the lens galaxies, we find that those
with late morphological types exhibit significant residuals after the
subtraction of the photometric model. It is therefore relatively easy
to remove the small number of spiral galaxies from the lens sample.
While there may be a few lens galaxies which cannot be definitively
typed using the above methods, signal-to-noise considerations present
yet another barrier for the inclusion of spirals. Because late-type
lenses tend to be both smaller and fainter than early-types, they are
far more likely to be dropped from the sample due to poor photometry
or contrast problems. Consequently, gravitational lensing offers a
mass-selected sample of early-type galaxies, with minimal
contamination from later morphological types.

Accurate photometry for gravitational lenses is a challenge, as
emission from the galaxy must be measured in the presence of multiple
compact quasar components. Space-based imaging is therefore essential
to properly decompose arcsecond-scale lens systems. To this end, the
CfA-Arizona Space Telescope Lens Survey
(CASTLES)\footnote{http://cfa-www.harvard.edu/castles/} has been using
the Hubble Space Telescope (HST) to observe known gravitational lenses
in the F555W=$V$, F814W=$I$ and F160W=$H$ bands. This is combined with
archival data obtained by other groups, which often includes
observations in additional WFPC2 and NICMOS filters. All data is
uniformly analyzed and model-fit according to the procedures described
in detail by Leh\'ar et al.\ (2000) and Kochanek et al.\ (2000). For
each lens galaxy, the intermediate axis effective radius ($r_e$) is
determined in the filter with the highest signal-to-noise ratio by
fitting a de Vaucouleurs profile convolved with an HST point spread
function model. The effective radius is assumed to be the same in all
bands. While local samples suggest that the effective radius mildly
decreases with increasing wavelength (e.g., Pahre, de Carvalho \&
Djorgovski et al.\ 1998b; Scodeggio 2001; Bernardi et al.\ 2001) due
to radial color gradients (de Vaucouleurs 1961), we cannot pursue this
effect in our intermediate redshift sample. Our signal-to-noise
degrades rapidly as one moves from the near-infrared to optical
filters, and a robust measurement of $r_e$ is rarely possible at $V$,
except for local lenses. In the cases where independent fits can be
performed, we find little evidence for significant wavelength
dependence in effective radius.\footnote{One exception is MG2016+112,
in which the galaxy has a much smaller scale radius at $H$ than at $I$
(e.g., Koopmans \& Treu 2002a). NICMOS $H$-band images show an Airy
ring about the compact galaxy core.} We therefore apply the effective
radius measured in the best filter to determine the enclosed
(``effective'') surface brightness $\mu_e$ in all filters. The total
magnitudes are then derived from the above quantities ($m = \mu_e - 5
\log r_e - 2.5 \log 2\pi$). These are corrected for Galactic
extinction using the formulae of Cardelli, Clayton \& Mathis (1989),
$R_V = 3.1$, and the appropriate $E(B-V)$ coefficients from Schlegel,
Finkbeiner \& Davis (1998).

The sample employed in this paper is similar to that analyzed by
Kochanek et al.\ (2000), which included 30 early-type lens
galaxies. As before, lens systems with poor photometry due to low
signal-to-noise (e.g., a faint, or even undetected galaxy) or high
contrast between the quasar images and galaxy emission are removed. We
exclude four systems from the Kochanek et al.\ (2000) sample. First,
lenses with large non-linear cluster perturbations complicate the
relationship between the image separation and galaxy properties, so we
reject Q0957+561 (Walsh, Carswell \& Weymann 1979) and RXJ0911+0551
(Bade et al.\ 1997). Second, we reject two systems in which neither
the lens nor source redshift is known -- B1127+385 (Koopmans et al.\
1999) and HST 12531--2914 (Ratnatunga et al.\ 1995) -- as galaxy
evolution is difficult to treat in this case. As before, Q2237+030
(Huchra et al.\ 1985) is retained because the lensing mass is the
ellipsoidal bulge of a spiral galaxy. Since the publication of
Kochanek et al.\ (2000), additional HST observations have filled in
gaps in the $V$--$I$--$H$ photometric coverage for a number of the
lenses, allowing for an improved determination of the lens galaxy
properties. The complete updated data set was re-analyzed and
model-fit for this paper. Recently measured redshifts have also been
incorporated into our analysis. Finally, two new lenses are added to
the sample: CTQ 414 (Morgan et al.\ 1999) and HS0818+1227 (Hagen \&
Reimers 2000). The total number of lenses used is therefore 28, and
their properties are summarized in Table~1.

\subsection{Spectrophotometric Models}

As a first analysis step, one must convert data from observed filters
into magnitudes in standard rest frame bands. We use the GISSEL96
version of the Bruzual \& Charlot (1993) spectral evolution models,
assuming $\Omega_m =0.3$, $\Omega_{\Lambda} = 0.7$, and $h =
0.65$. Given a star-formation history (typically modeled as a
starburst of some duration $\tau$ beginning at redshift $z_f$), an
initial mass function (IMF) and a metallicity $Z$, the models predict
a spectral energy distribution (SED) as a function of redshift. The
SED can then be convolved with filter transmission curves to compute
synthetic colors. We take as a fiducial model an instantaneous ($\tau
= 0$) burst at $z_f=3$ with solar metallicity $Z=Z_{\odot}$ and a
Salpeter (1955) IMF.  The assumed metallicity is broadly consistent
with that of early-type galaxies at $0.3 < z < 0.9$ (Ferreras, Charlot
\& Silk 1999), and the assumed formation redshift has little effect on
our results (see below).  We use filter curves appropriate for HST
filters (available from the technical archives of STScI), with
zero-points from Holtzman et al.\ (1995).

The spectrophotometric model allows us to interpolate observed bands
to a given rest frame band. For example, the observed $V$ and $I$
filters bracket rest frame $B$ over most of the redshift range spanned
by the lens galaxies. The interpolation uses synthetic ``colors''
$C(X,Y) \equiv m_{mod,X} - m_{mod,Y}$ between rest frame magnitudes in
filter $X$ and directly measurable magnitudes in filter $Y$,
calculated from the model SED for a galaxy at the appropriate
redshift. The rest frame magnitude $m_X$ is then derived from the
observed HST magnitudes $m_{obs,Y}$ and colors:
\begin{equation}
m_{X} = \frac{\sum_Y[m_{obs,Y}+C(X,Y)]/(\delta m_{obs,Y})^2}{\sum_Y
1/(\delta m_{obs,Y})^2} \ ,
\end{equation}
where the sum is taken over all filters in which the galaxy has been
observed.  We find that the scatter in $m_{obs,Y}+C(X,Y)$ amongst the
various filters tends to be small ($\la 0.15$ mag), indicating that
most lens galaxy colors are consistent with those of our standard
spectrophotometric model. The specific choice of evolution model has
little effect on the derived magnitudes, as only the shape of the
spectrum is relevant for the interpolation, and this does not change
dramatically among reasonable models. For example, averaging over a
broad range of stellar models ($1 < z_f < 5$, $0.4 Z_{\odot} < Z < 2.5
Z_{\odot}$) typically results in a scatter of $\la 0.1$ mag in the
interpolated magnitudes, even if we keep models which fit the observed
colors poorly.

Nominal errors on the derived magnitudes tend to be small ($\la 0.1$
mag), as most lens galaxies in our analysis sample have good
photometry in at least one filter. However, to account for the
variation in the rest frame magnitudes due to the ensemble of
spectrophotometric models ($\sim 0.1$ mag) and the rms in
filter-to-filter estimates $m_{obs,Y}+C(X,Y)$ due to slightly
discrepant colors ($\sim 0.15$ mag), we simply set a uniform
uncertainty of $\delta M = 0.2$ mag on the derived absolute magnitudes
(at fixed galaxy redshift). The value of the assumed photometric
errors has little effect on either the derived evolution rates or
their error bars, as we later rescale all uncertainties to reflect the
observed scatter in the galaxy ensemble (\S 3.2). For a lens galaxy
with an estimated redshift and a corresponding redshift uncertainty,
the correlation between the absolute magnitude and redshift is taken
into account (\S 2.5).

\subsection{Estimating Velocity Dispersions from Image Separations}

While the lensed image separation is measured directly, the critical
radius that sets the splitting scale for the lens must be
inferred. Fortunately, the critical radius is closely related to the
image separation, and is essentially independent of the chosen mass
model (e.g., Schneider et al.\ 1992). We therefore crudely fit each
lens system with a singular isothermal sphere (SIS) in an external
shear field. The standard image separation parameter is then defined
as $\Delta \theta \equiv 2 b$, where $b$ is the Einstein radius of the
SIS. This method is preferable to simply using the maximum separation
between lensed images or twice their mean distance from the galaxy
center. Such estimators can be strongly affected by large shears and
ellipticities, and are inappropriate for certain types of quad
lenses. Note that the scatter in $\Delta \theta$ introduced by
different mass models or fitting techniques is small ($\la 2\%$), and
we can safely ignore it in our analysis.\footnote{ Because the
velocity dispersion $\sigma \propto (\Delta \theta)^{1/2}$, even a
large error of $5\%$ in $\Delta \theta$ would lead to an error of only
$\sim 0.01$ in $\log \sigma$, which is the relevant quantity for
evolution studies (\S 3.1).}

It has been repeatedly demonstrated that the mass distributions of
early-type gravitational lens galaxies are very close to isothermal
(Kochanek 1995; Cohn et al.\ 2001; Mu\~noz et al.\ 2001; Rusin et al.\
2002; Treu \& Koopmans 2002a; Koopmans \& Treu 2002b), consistent with
measurements from stellar dynamics (Rix et al.\ 1997; Gerhard et al.\
2001) and X-ray halos (Fabbiano 1989). An SIS produces an image
separation of $\Delta \theta = 2 b = 8 \pi (D_{ds}/D_{s})
(\sigma_{DM}/c)^2$, where $\sigma_{DM}$ is the dark matter velocity
dispersion, and $D_{ds}$ and $D_s$ are angular diameter distances from
the lens to the source, and from the observer to the source,
respectively. Because the mean image separation produced by a given
lensing galaxy depends on the lens and source redshifts,\footnote{The
splitting is also weakly cosmology-dependent, but we assume a flat
$\Omega_m = 0.3$ universe throughout this paper.}  the redshift
dependence must be removed in order to properly interpret the observed
image separation. This is accomplished by defining a reduced image
separation $\Delta \theta_{red} \equiv (\Delta \theta / \Delta
\theta_*)(D_s / D_{ds})$, where $\Delta \theta_* = 8 \pi
(\sigma_{DM*}/c)^2 \simeq 2\farcs91$, assuming a characteristic
velocity dispersion of $\sigma_{DM*} \simeq 225\kms$ for an $L_*$
galaxy (e.g., Kochanek 1994, 1996). Note that the parameter $\Delta
\theta_{red}$ represents a physical property of the lensing
galaxy. The values of $\Delta \theta_{red}$ for each lens are listed
in Table~1.

For an isothermal model, $\sigma_{DM}$ is close to the ``central''
stellar velocity dispersion $\sigma_c$ (e.g., Kochanek 1994; KT),
which is often defined within a standard aperture of 3\farcs4
(diameter) at the distance of the Coma cluster (e.g., Jorgensen et
al.\ 1996). For a given mass and luminosity distribution, $\sigma_c$
can be calculated by solving the Jeans equation (e.g., Binney \&
Tremaine 1987). The overall mass distribution is assumed to be
isothermal, while the luminosity distribution is modeled by a
Hernquist (1990) profile with characteristic radius $a = 0.551
r_e$. We compute $\sigma_c/\sigma_{DM}$, integrated inside the ``Coma
aperture'', assuming isotropic orbits (see, e.g., Kochanek 1993b for
illustrative plots). The choice of anisotropy parameter has little
effect on any of the fit results presented below. Furthermore, there
is little difference between using the Coma aperture, a standard
aperture equal to $r_e / 8$, or simply using $\sigma_{DM}$, as
demonstrated in \S 3.2.

The distribution of dark matter velocity dispersions ($\sigma_{DM}$)
for the gravitational lens sample has a median of $226\kms$, a mean of
$233\kms$, and a standard deviation of $50\kms$. The estimated stellar
velocity dispersions ($\sigma_c$) inside the Coma aperture have a
median of $215\kms$, a mean of $226\kms$, and a standard deviation of
$55\kms$. Consequently, our analysis will probe the evolution of
early-type galaxies close to $L_*$.

\subsection{Unmeasured Lens and Source Redshifts}

Both lens and source redshifts are needed to properly interpret galaxy
magnitudes and image separations.  First, the lens redshift is
required to determine the absolute magnitude and consider its
evolution. Second, each redshift is needed to relate the splitting
scale to physical properties of the lensing galaxy (\S
2.3). Unfortunately, spectroscopic follow-up continues to be a major
bottleneck for strong gravitational lensing. Approximately one third
of the 28 lenses meeting our selection criteria have incomplete
redshift information, but with careful consideration, these systems
can be included in the analysis.

For systems with only a measured source redshift $z_s$, we use the FP
redshift estimation technique outlined in Kochanek et al.\
(2000). Specifically, we determine the lens redshift $z_d$ at which
the galaxy must reside in order to most nearly evolve to the local FP:
$\log R_e = \alpha \log \sigma_c + \beta \langle SB_e \rangle +
\gamma$, where $R_e$ is the effective radius in physical units,
$\sigma_c$ is the central stellar velocity dispersion, and $\langle
SB_e \rangle$ is the mean absolute surface brightness within
$R_e$.\footnote{Our convention is such that the observed (apparent)
quantities are denoted as $r_e$ and $\mu_e$, while the physical
(absolute) quantities are denoted as $R_e$ and $\langle SB_e
\rangle$.} Our implementation differs from Kochanek et al.\ (2000) in
that we use the rest frame $B$-band FP (Bender et al.\ 1998; see also
\S 3.1). The extracted redshift varies systematically with the assumed
evolution model, since the comparison must be made at $z=0$. Redshift
estimation is therefore the one aspect of our analysis in which
choosing a particular SED model can bias later measurements of the
evolution rate. To avoid introducing any significant bias, we consider
a broad range of evolution rates, $ -0.8 < d\log (M/L)_B / dz < -0.2$,
which more than spans the reasonable range of star-formation epochs
($z_f \ga 1$). The mean FP redshift is used as the redshift estimate,
and the rms as the uncertainty. The FP redshifts of intrinsically
high-redshift lens galaxies are very uncertain, as the effect of the
different evolution models is large. Low-redshift galaxies, however,
should have robust redshift estimates. Finally, for systems with a
measured $z_d$ but no $z_s$, we simply assume $z_s = 2.0 \pm 1.0$,
consistent with the typical range for lensed sources.

\subsection{Combining Uncertainties}

Propagating the uncertainties in observed and derived quantities
becomes increasingly complex as spectroscopic redshift information is
removed. For lenses with a measured $z_d$ and $z_s$, our methods yield
effectively no uncertainty on $\Delta \theta_{red}$, and hence on the
inferred $\sigma_{DM}$ and $\sigma_c$. The measurement uncertainty in
the photometric quantity $\log r_{e} - \beta \mu_{e}$ (where $\beta
\simeq 0.32$; see \S 3.1) is negligible, since fitting errors in the
individual quantities are highly correlated and nearly
cancel. However, because there is an independent error introduced by
converting between the observed magnitudes and rest frame magnitudes,
the quantity entering the FP ($\xi \equiv \log R_{e} - \beta \langle
SB_e \rangle$) has an error $\delta \xi \simeq \beta \delta M \simeq
0.06$.

For lenses with a measured $z_d$ and an assumed $z_s=2.0 \pm 1.0$, the
photometric errors are the same as described above, but there are
uncertainties in $\Delta \theta_{red}$ and $\sigma_c$ that must be
determined numerically. This is accomplished by drawing 10000 values
from a Gaussian distribution representing the source redshift
($\langle z_s \rangle =2.0$, width $1.0$), and calculating the derived
quantities. Only trials with $1 \leq z_s \leq 5$ are accepted, as
virtually all known lensed sources reside within this redshift
range. The rms scatters of $\log \Delta \theta_{red}$ and $\log
\sigma_c$ about their $z_s = 2.0$ values are used as their
uncertainties.

For lenses with a measured $z_s$ and an FP-estimated $z_d$, the
uncertainty in $z_d$ induces correlated uncertainties in the absolute
magnitude, FP combination $\xi$, $\Delta \theta_{red}$ and
$\sigma_c$. To account for this, 10000 trials are drawn from a
Gaussian distribution representing the lens redshift ($\langle z_d
\rangle =z_{FP}$, width $\delta z_{FP}$), and all derived parameters
are calculated and stored for later use. The quantities entering the
fits to the FP or Faber-Jackson relation (\S 3 and \S 4) contain
various combinations of these correlated inputs. We construct
realistic tolerances on individual data points by calculating the rms
scatter of the quantity entering the fits, using the 10000 sets of
correlated parameters. 

Finally, the error in $r_e$ induces an uncertainty in $\sigma_c$, as
an uncertain fraction of the radial luminosity profile is probed by
the fixed aperture $r_{ap}$. We estimated the uncertainty by drawing
10000 trials from a Gaussian distribution representing the effective
radius (mean $\langle \log r_e \rangle$, width $\delta \log r_e$), and
calculating $\log \sigma_c$ for each. We find that the uncertainty in
$\log \sigma_c$ due to $\delta \log r_e$ is negligible, as the
dynamical correction is a slowly-varying function of $\log
(r_{ap}/r_e)$ and the errors on $\log r_e$ are small ($\la 0.1$) for
most of the lens galaxies in our sample (Table~1).

\section{Measuring Evolution Using the Fundamental Plane}

\subsection{Background}

The fundamental plane of early-type galaxies (Djorgovski \& Davis
1987; Dressler et al.\ 1987) describes the correlation among physical
effective radius $R_e$, central stellar velocity dispersion
$\sigma_c$, and mean absolute surface brightness $\langle SB_e\rangle$:
\begin{equation}
\log (R_e / \kpc) = \alpha \log (\sigma_c / \kms) + \beta (\langle
SB_e \rangle / {\rm mag \, arcsec^{-2}}) + \gamma \ .
\end{equation} 
The parameters $\alpha$ and $\gamma$ depend on wavelength (see Pahre
et al.\ 1998b for theoretical motivations), but $\beta \simeq 0.32$
does not (e.g., Scodeggio et al.\ 1998; Pahre et al.\ 1998a). There is
not yet any convincing evidence of differences between the field and
cluster FP (e.g., Kochanek et al.\ 2000; Treu et al.\ 2001; van Dokkum
et al.\ 2001), but such investigations are still in their formative
stages. Throughout this section we will assume that at a given
wavelength, both field and cluster galaxies fall on an FP with
identical slopes.

In the context of evolution studies, the FP allows us to predict the
surface brightness that a galaxy would have at $z=0$, given its
velocity dispersion and effective radius. As in all previous FP
evolution analyses, we assume that these structural parameters do not
evolve. The difference between the observed surface brightness and the
predicted $z=0$ value is then directly related to the luminosity
evolution: $\Delta \log L = -0.4 \Delta \langle SB_e \rangle = \Delta
\gamma/ (2.5 \beta)$. Consequently, while we phrase our evolution
results in terms of ``mass-to-light'' ratios, a ``mass'' never need
explicitly enter. We are simply using an empirical relation among
observables (in this case, the FP) to predict the surface brightness
at $z=0$, which is compared to the measured value at $z_d$ to yield
the evolution rate. However, if one defines an effective mass $\propto
\sigma_c^2 R_e/G$, the effective mass-to-light ratio can be written in
terms of FP parameters and observables (see, e.g., Treu et al.\ 2001
for a complete discussion):
\begin{equation}
\log (M/L) \propto \left( {10\beta - 2\alpha \over 5 \beta} \right)
\log \sigma_c + \left( {2-5\beta \over 5 \beta} \right) \log R_e -
{\gamma \over 2.5 \beta} \ .
\end{equation}
If $\alpha$ and $\beta$ are constant, then the evolution rate is
determined solely by the redshift-dependent intercept of the FP, $d
\log (M/L) / dz = - (1/2.5 \beta)d\gamma / dz $, as described
above. We will assume that the FP slopes are independent of redshift,
as has been done in recent studies of early-type field galaxies (van
Dokkum et al.\ 2001; Treu et al.\ 2001, 2002; KT). The assumption is
well motivated. Currently there is no observational evidence for
redshift evolution in $\beta$ (e.g., Kelson et al.\ 2000), while there
is some disagreement regarding $\alpha$. Pahre et al.\ (2001) have
claimed strong $\alpha$ evolution at $K$ band, but Kelson et al.\
(2000) find little evolution at $V$ band. A redshift-dependent
$\alpha$ complicates the study of luminosity evolution because
galaxies with different velocity dispersions would evolve at different
rates. However, these effects only become important when a wide range
of velocity dispersions are present in the galaxy sample. Like most
samples of early-type field galaxies (e.g., Treu et al.\ 2001), the
lenses span a rather narrow range in velocity dispersion (only $\sim
0.1$ dex about $\langle \sigma_c \rangle \sim 225\kms$; see \S
2.3). Because we are exploring luminosity evolution over a small range
of mass, we may safely ignore $d \alpha / dz$ and parametrize the
problem solely in terms of $d \gamma / dz$. We will therefore robustly
constrain $d \log (M/L) / dz$ at $200\kms \la \sigma_c \la 250\kms$,
regardless of any potential evolution in $\alpha$.

\subsection{Calculations and Results}

We now investigate the luminosity evolution at rest frame $B$ band,
using the redshift-dependent intercept of the FP. The parameters of
the local $B$-band FP are set to $\alpha = 1.25$, $\beta = 0.32$ and
$\gamma_0 = -8.895 - \log(h/0.5) \pm 0.01$, based on observations of
the Coma cluster (Bender et al.\ 1998).\footnote{Following Treu et al.\
(2001, 2002), we assume that these parameters describe a hypothetical
$z=0$ sample, even though they are derived at $z\sim 0.02$.} We
reiterate that no significant differences in the slope or intercept
have been detected between field and cluster ellipticals (e.g.,
Kochanek et al.\ 2000; van Dokkum et al.\ 2001). Based on the
arguments of \S 3.1, we assume no redshift dependence of $\alpha$ and
$\beta$, and parameterize the evolution solely in terms of the
intercept ($\gamma$). For each lens galaxy $i$, we construct
\begin{equation}
\gamma_i = \log R_{e,i} - \alpha \log \sigma_{c,i} - \beta \langle SB_e \rangle_i
\end{equation}
from the available data, and compare this to the assumed present-day
value ($\gamma_0$). The corresponding mass-to-light ratio offset is
then modeled as $\Delta \log (M/L)_i = -(\gamma_i - \gamma_0)/(2.5
\beta) = a z$, where $a \equiv d \log (M/L)/ dz \equiv - (1/2.5\beta)
d\gamma / dz$. Note that for passively evolving stellar populations
formed at $z_f > 1.5$, the luminosity evolution at $z<1$ is nearly
linear in redshift (Fig.~1; see also van Dokkum \& Franx 1996). The
evolution rate is then optimized via the goodness-of-fit parameter
\begin{equation}
\chi_{FP}^2 = \sum_{i=1}^{N_{gal}}{
[\Delta \log (M/L)_i - a z_{d,i}]^2 \over \delta_i^2} \ ,
\end{equation}
where $\delta_i^2 = \delta_{scale}^2 \delta^2[\Delta \log (M/L)_i - a
z_{d,i} ]$, and $\delta_{scale}$ is an overall scale factor that is used to
estimate the parameter uncertainties (see below). If the lens redshift
is known, $\delta_i^2 = \delta_{scale}^2 \{(1/2.5 \beta)^2 [(\delta
\xi)^2 + \alpha^2 (\delta \log \sigma_{c,i})^2]\}$. If the lens redshift
is estimated, the correlation among all quantities is taken into
account by constructing $\delta_i^2$ via Monte Carlo (\S 2.5). 

Uncertainties on individual fit parameters are estimated from the
$\chi^2$, bootstrap resampling and jackknife methods (e.g., Press et
al.\ 1992). For the $\chi^2$ method, we uniformly rescale the input
errors by setting $\delta_{scale}$ so that the best-fit model has
$\chi^2 = N_{DOF}$, the number of degrees of freedom. This does not
alter the optimized parameters, but does allow us to compensate for
possibly under-estimated errors in our data and relate the
uncertainties to the observed scatter. We then calculate the
one-dimensional parameter ranges with rescaled $\Delta \chi^2 \leq 1$
($1\sigma$). To derive the bootstrap errors we generate 10000
resamplings of the data set, and determine the best-fit model
parameters for each. It is found that all three techniques give
consistent estimates of the error bars, and we quote the $\chi^2$
errors as our parameter uncertainties.

The rest frame $B$-band mass-to-light ratio offsets (versus the
FP-predicted present-day values) are plotted as a function of redshift
in Fig.~2. A linear decrease in $\log (M/L)_B$ with redshift is clear,
indicating that lens galaxies were brighter in the past. From the
above analysis, we obtain an evolution rate among early-type lens
galaxies of $d\log (M/L)_B/dz = -0.56 \pm 0.04$ ($1 \sigma$). We find
no significant difference between low and high-redshift subsamples of
gravitational lenses: $d \log (M/L)_B/dz = -0.64 \pm 0.09$ for the 14
galaxies with $z_d \leq 0.5$, versus $d \log (M/L)_B/dz = -0.55 \pm
0.05$ for the 14 galaxies with $z_d > 0.5$.  Restricting the sample to
only those 22 lens systems with spectroscopic galaxy redshifts results
in $d\log (M/L)_B/dz = -0.58 \pm 0.05$, so there is little bias or
extra scatter introduced by our redshift estimation scheme. Using the
high measured velocity dispersion of PG1115+080 (Tonry 1998) instead
of the isothermal estimate moves the galaxy off the FP, as discussed
by Treu \& Koopmans (2002b), but does not affect the derived evolution
rate for the sample as a whole. Furthermore, while we have placed no
uncertainties on the estimated velocity dispersions, including errors
comparable to typical measurement errors (e.g., Treu et al.\ 2001) has
little effect on any results. There is, however, a small systematic
effect related to the choice of aperture in which the stellar velocity
dispersion is estimated. Our standard aperture is equal to 3\farcs4
(diameter) at the distance of Coma. If we choose apertures with radius
$r_e/8$, the evolution rate is $d\log (M/L)_B/dz = -0.50 \pm
0.04$. Simply setting $\sigma_c = \sigma_{DM}$ yields $d\log
(M/L)_B/dz = -0.54 \pm 0.04$.  

The evolution rate of mass-selected early-type field galaxies is
slower than that of the luminosity-selected sample of Treu et al.\
(2001, 2002), who find $d\log (M/L)_B/dz = -0.72^{+0.11}_{-0.16}$. The
cluster ($d \log (M/L)_B / dz = -0.49 \pm 0.05$) and field ($d \log
(M/L)_B / dz = -0.59 \pm 0.15$) samples of van Dokkum et al. (1998,
2001) more closely agree with our results.\footnote{Note that the
analysis techniques of van Dokkum et al.\ (1998, 2001) and Treu et
al.\ (2001, 2002) differ slightly. Treu et al.\ compare the inferred
FP intercepts with the local intercept derived from the Coma cluster
(Bender et al.\ 1998). van Dokkum et al.\ appear to re-fit the local
intercept.}  The uncertainty in our evolution rate is smaller than
those of previous samples, primarily because the lenses extend deeper
in redshift ($z \sim 1$), providing a longer lever-arm with which to
trace the evolution.  The unweighted rms scatter in $\Delta \log
(M/L)_B$ about our best-fit model is $0.13$, which corresponds to a
scatter of $0.11$ in the FP, slightly larger than the values found in
local luminosity-selected samples (e.g., Jorgensen et al.\ 1996; Pahre
et al.\ 1998a; Bernardi et al.\ 2001).  The additional scatter may be
due to our velocity dispersion estimation technique, or the broad
redshift range of our sample. The scale factor must be set at
$\delta_{scale} \sim 1.5$ for our error bars to properly reflect the
observed scatter. Because the scatter is larger than can be accounted
for by measurement errors, some fraction of the scatter may be
intrinsic to the FP (Lucy, Bower \& Ellis 1991; Jorgensen et al.\
1996).

We can compare our results for MG2016+112 and 0047--281 with those
derived by KT, who have combined archival HST data with direct
measurements of the lens velocity dispersion. For MG2016+112 we find
$\Delta \log (M/L)_B = -0.62 \pm 0.08$ between $z=0$ and $z=1.00$,
versus $-0.62 \pm 0.08$ (Koopmans \& Treu 2002a). For 0047--281 we
find $\Delta \log (M/L)_B = -0.41 \pm 0.08$ between $z=0$ and
$z=0.49$, versus $-0.37 \pm 0.06$ (Koopmans \& Treu 2002b). The
agreement is striking, considering our independent analyses of the HST
data, and the use of independent codes and assumptions for converting
between observed and rest frame magnitudes. Furthermore, using
estimated stellar velocity dispersions instead of measured values has
little effect. We therefore believe that our estimate of the evolution
rate for the entire lens sample is robust. Finally, it is interesting
to point out that the evolution rates derived from the two KT lenses
alone are systematically faster than the lens sample as a whole, as
both galaxies fall below the evolutionary trend lines in Fig.~2.

Next we simultaneously fit for both the evolution rate and the
present-day FP intercept. This technique offers a more realistic
estimate of the observed redshift trend, as the fit is no longer
forced through a fixed zero-point. The extracted evolution rate should
also be less biased by possible systematic differences between the FP
parameters for luminosity-selected and mass-selected galaxies, as well
as cluster and field populations. In addition to removing the explicit
comparison to a luminosity-selected local intercept derived in a
cluster environment, the new fit also minimizes the effect of small
changes in the slopes. For example, there is a degeneracy between
$\alpha$ and the intercept, such that $\gamma_0 \propto - \langle \log
\sigma_c \rangle \alpha$. Fixing $\alpha$ at different values (that
are close to reality) will systematically change $\gamma_0$, but only
minimally affect $d \gamma / dz$, from which the evolution rate is
derived. Moreover, re-fitting the present-day intercept decreases any
systematic effects related to our velocity dispersion estimation
technique. For example, if lens galaxies have mass profiles that are
slightly steeper than isothermal, then the isothermal model will
systematically under-estimate the stellar velocity dispersion, and
vice versa. Fractionally changing $\sigma_c$ will require an
adjustment of the present-day FP intercept, but will not affect the
redshift trend. Consequently, simultaneously re-fitting the local
intercept should lead to an estimate of the evolution rate that is
relatively immune to systematics.

Constraints on the $B$-band evolution rate and local FP intercept are
plotted in Fig.~3. The two quantities are highly correlated. The
best-fit evolution rate (assuming $\alpha = 1.25$, $\beta = 0.32$) is
$d \log (M/L)_B / dz = - 0.54 \pm 0.09$, slower but still consistent
with the value derived by fixing the local FP intercept. The best-fit
intercept is $\gamma_0 = -8.88 - \log (h/0.5) \pm 0.04$, consistent
with the value of $\gamma_0 = -8.895 - \log (h/0.5) \pm 0.01$ derived
from Coma (Bender et al.\ 1998).  These results suggest that by
assuming luminosity-selected cluster FP parameters, one does not
significantly bias the evolution rate derived for a mass-selected
field galaxy sample.  As described above, the $\alpha$-dependence of
the derived evolution rate is small if the intercept is allowed to
vary. For example, we find $d \log (M/L)_B / dz = - 0.56 \pm 0.09$ and
$d \log (M/L)_B / dz = - 0.51 \pm 0.09$ when we fix $\alpha = 1.15$
and $1.35$, respectively. Finally, if we set $\sigma_c = \sigma_{DM}$,
we find $d \log (M/L)_B / dz = - 0.58 \pm 0.09$. Identical evolution
rates are achieved if we use velocity dispersions estimated inside any
fixed fraction of the effective radius, as our simple dynamical model
implies that $\sigma_c / \sigma_{DM}$ is the same for all lens
galaxies in this case.

Finally, we consider the evolution of the FP mass-to-light ratio at
longer rest frame wavelengths. To accomplish this, we simply
interpolate our observed magnitudes to the rest frame equivalents of
the HST $V$, $I$ and $H$ bands and repeat the analysis. Rest frame $V$
and $I$ are bracketed by the observed $V$, $I$ and $H$ filters out to
$z = 1$. Rest frame $H$ is an extrapolation of our data
set. Early-type galaxies in local clusters have been investigated in
nearby bands, yielding estimates of the slopes $\alpha$ and
$\beta$. Unfortunately, many of these analyses lack robust
determinations of $\gamma$, so we will again include the intercept as
part of the fit. We fix ($\alpha$, $\beta$) to the representative Coma
values measured by Scodeggio et al.\ (1998): ($1.35$, $0.35$) at $V$,
($1.70$, $0.34$) at $I$, and ($1.66$, $0.34$) at $H$. The derived
evolution rates will be only weakly sensitive to the particular value
of $\alpha$, as described above. We find $d \log(M/L)/dz = -0.46 \pm
0.09$ at $V$, $-0.31 \pm 0.10$ at $I$, and $-0.24 \pm 0.10$ at
$H$. This confirms the expected trend of slower evolution rates at
longer rest frame wavelengths. However, the above values are not
independent, as they are based on the same set of observed
magnitudes. Note that the evolution rate derived from the FP slows
with wavelength more dramatically than one might predict from the
model SED. Hence, the band-to-band evolution rates suggest rather
large rest frame color evolution. The result is deceptive, however, as
a consistent rest frame color is never constructed in our analysis. We
trace the seemingly anomalous color evolution to a combination of
$\alpha$ increasing with wavelength, and the mean velocity dispersion
increasing with redshift (see \S 4.2 and Fig.~7 for more details).

\section{Aperture Mass-to-Light Ratios and the Faber-Jackson Relation}

\subsection{Background}

Gravitational lenses directly measure the projected mass inside the
aperture defined by the Einstein radius $b = \Delta \theta/2$, 
\begin{equation}
M_{ap} = \pi \Sigma_{cr} D_d^2 \left( {\Delta \theta \over 2}
\right)^2 \propto {D_d D_{ds} \over D_s} (\Delta \theta_{red})^2 \ ,
\end{equation}
where $\Sigma_{cr} = c^2 D_s/ 4\pi G D_{ds} D_d$ is the critical
surface mass density (Schneider et al.\ 1992). This aperture mass
measure is a powerful tool, as it is very precisely determined and
does not depend on the radial mass profile. Because aperture
luminosities can be easily derived from the total magnitudes,
gravitational lensing offers a unique mass-to-light ratio for use in
evolution studies (Keeton et al.\ 1998).

While the aperture technique precisely determines the mass-to-light
ratio inside a particular radius for a particular galaxy, several
corrections are needed to investigate luminosity evolution using a
sample of lenses at various redshifts. Consider a galaxy with a fixed
mass distribution and luminosity, which lenses a background source at
redshift $z_s$. As the galaxy is moved to different redshifts $z_d$,
the aperture radius $b$ defined by the lensed images changes, as does
the enclosed mass (due to both $b$ and $\Sigma_{cr}$). The fraction of
the luminosity enclosed by the aperture depends on $r_e/b$, where
$r_e$ is the angular effective radius, so this will also change. A
scheme for removing the redshift effects must therefore satisfy one
essential criterion: a non-evolving galaxy must yield the same
mass-to-light ratio regardless of where it is placed in
redshift. Because the mass and light will in general have different
profiles, it is necessary to correct each independently. To handle an
ensemble of galaxies with different masses, the mass-to-light ratio
must also be normalized to a common corrected mass scale.

From Eq.~6, a corrected redshift-independent aperture mass is
$M_{ap}^{corr} \equiv M_{ap} D_s / D_d D_{ds} \propto (\Delta
\theta_{red})^2$. The aperture luminosity must be corrected to a fixed
value of $r_e/b$ (any constant fraction of the effective radius), and
hence, it will be proportional to the total luminosity: $L_{ap}^{corr}
\propto L$. Since we are dealing with an ensemble of galaxies, we
include a term (with proportionality constant $a_2$) to normalize all
galaxies to a common corrected mass scale. 
We therefore define a corrected aperture mass-to-light ratio
$\Upsilon$ as follows:
\begin{eqnarray}
  \log\Upsilon &\equiv& \log(M/L)_{ap}^{corr} + a_2 \log M_{ap}^{corr} +
    \mbox{const}\,, \\
  &=& 2(1+a_2)\log\Delta\theta_{red} + 0.4 M + \mbox{const}\,,
\end{eqnarray}
where in the second line we have converted luminosity to absolute
magnitude using $M = -2.5\log L + \mbox{const}$.  We then postulate
that the mass-to-light ratio evolves with the simple form
\begin{equation}
  \log\Upsilon = \Upsilon_0 + \gamma_{EV} z\,, 
\end{equation}
where $\Upsilon_0$ is a zero-point and $\gamma_{EV}$ is the rate of
evolution. Combining Eq.~8 and 9, we can write
\begin{equation}
  M = M_{*0} + 2.5 \gamma_{EV} z - 1.25 \gamma_{FJ} \log\Delta\theta_{red}\,,
\end{equation}
where $\gamma_{FJ} = 4(1+a_2)$ and $M_{*0}$ is the present-day
characteristic magnitude for a galaxy with $\sigma_{DM*}$. This is the
lensing Faber-Jackson relation, a relation between absolute magnitude
and image separation, where the image separation serves as a proxy for
the velocity dispersion that appears in the standard Faber-Jackson
relation (Keeton et al.\ 1998).  The coefficient $\gamma_{FJ}$ is
chosen so that we have the scaling $L \propto
(\Delta\theta_{red})^{\gamma_{FJ}/2} \propto
\sigma_{DM}^{\gamma_{FJ}}$. We fit to Eq.~10 directly, rather than
converting the reduced image separations into stellar velocity
dispersions, because the relation between luminosity and splitting
scale is really the fundamental relation for gravitational lens
galaxies (as illustrated by the corrected aperture mass-to-light
ratios).  However, the isothermality of the lens population, which
implies $\sigma_{DM} \sim \sigma_c$, does allow for a straightforward
comparison between our FJ slope and those derived for the general
population of early-type galaxies.

As with our FP analysis, we are not measuring a true mass-to-light
ratio, but rather using an empirical relation to normalize the galaxy
luminosities to a fixed scale ($\Delta
\theta_{red})^{\gamma_{FJ}/2}$. By minimizing the scatter between the
predicted and observed magnitudes, we can use the FJ relation to
simultaneously constrain $d\log(M/L)/dz$, $M_{*0}$, and
$\gamma_{FJ}$. Finally, note that the derived value of $M_{*0}$ is
degenerate with $\Delta \theta_*$. We measure only the composite
parameter $M_{*0} + 1.25 \gamma_{FJ} \log (\Delta
\theta_*/2\farcs91)$.

\subsection{Calculations and Results}

The scatter of the lensing FJ relation is minimized via the
goodness-of-fit function
\begin{equation}
\chi_{FJ}^2 = \sum_{i=1}^{N_{gal}} {
(M_{mod,i} - M_{i})^2 \over \delta_i^2} \ ,
\end{equation}
where $\delta_i^2 = \delta_{scale}^2 \delta^2 (M_{mod,i} - M_{i})$. If
the lens redshift has been spectroscopically measured, $\delta_i^2 =
\delta_{scale}^2 [(\delta M_{i})^2 + 1.25^2 \gamma_{FJ}^2 (\delta
\log \Delta \theta_{red,i} )^2]$. If the lens redshift has been
estimated, the correlation among all quantities is taken into account
by constructing $\delta_i^2$ via Monte Carlo (\S 2.5).

At rest frame $B$ band, we find a present-day characteristic magnitude
of $M_{*0} = -19.70 + 5\log h \pm 0.29$ (for $\sigma_{DM*} = 225
\kms$), a Faber-Jackson slope of $\gamma_{FJ} = 3.29 \pm 0.58$, and an
evolution rate of $d \log (M/L)_B/dz = -0.41 \pm 0.21$. Fig.~4 shows
the input $B$-band magnitudes versus the predicted magnitudes for the
best-fit model parameters.  The lensing FJ relation forms a coherent
correlation over $\sim 4$ mag, with an unweighted rms scatter of
$0.53$ mag. We define a normalized luminosity $\log L_{B,norm} = -0.4
(M - M_{*0} + 1.25 \gamma_{FJ} \log \Delta \theta_{red})$ to place all
galaxies on a common splitting scale. The effective FJ mass-to-light
ratio $\log (M/L)_{B} \equiv -\log L_{B,norm}$ is plotted in Fig.~5.
While there is significantly more scatter than in the FP mass-to-light
ratios (due to the intrinsically larger scatter of the FJ relation;
e.g., Pahre et al.\ 1998a), a redshift trend is clear. Confidence
regions among pairs of fit parameters are plotted in Fig.~6. There are
correlations among all three quantities. Of particular interest is the
correlation between the evolution rate and FJ slope. This is the
result of the increase in the mean $\log \Delta \theta_{red}$ with
redshift (Fig.~7). Such a trend is expected from the angular splitting
selection function for lenses, as more massive galaxies are needed to
produce a fixed angular image separation as the galaxy redshift is
increased. However, the absence of low-redshift lens galaxies with
large $\log \Delta \theta_{red}$ is puzzling, and this exacerbates the
expected trend. It is easy to see how the observed correlation of the
mean $\log \Delta \theta_{red}$ with redshift results in correlated
values of $d \log (M/L)_B/ dz$ and $\gamma_{FJ}$. For example, one can
make high-redshift galaxies fainter with either a slower evolution
rate (so that they are not much brighter than present-day), or a
smaller $\gamma_{FJ}$ (so that the luminosity does not increase
strongly with the splitting scale, which is characteristically large
among high-redshift lenses).

Fits are also performed for the rest frame F555W=$V$, F814W=$I$ and
F160W=$H$ bands. The results are summarized in Table~2. Nearly
identical values of $\gamma_{FJ}$ are found in all bands, and the
evolution rates decrease with increasing wavelength. As with the FP
fits in multiple rest frame bands, the estimates are not independent.

The FJ evolution rate is consistent with the FP values, but has a much
larger uncertainty. The estimated FJ slope for lenses ($\gamma_{FJ} =
3.29 \pm 0.58$ at $B$ band) is broadly compatible with recent
measurements ($\gamma_{FJ} \simeq 4.0$) in the optical (Forbes \&
Ponman 1999; Bernardi et al.\ 2001) and near-infrared (Pahre et al.\
1998a). Our scatter of $0.53$ mag about the mean FJ relation is
somewhat smaller than found in the above analyses. Still, we must set
$\delta_{scale} \ga 2$ for the error bars to reflect the observed
scatter, indicating that much of the scatter is intrinsic to the FJ
relation. The $B$-band value of $M_{*0} = -19.70 + 5\log h \pm 0.29$
for gravitational lens galaxies is consistent with luminosity function
measurements (e.g., Efstathiou, Ellis \& Peterson 1988; Madgwick et
al.\ 2002). Finally, our estimates of $M_{*0}$ and $\gamma_{FJ}$ are
consistent with those of Keeton et al.\ (1998), who performed an FJ
analysis on a small sample of gravitational lens galaxies. We note
that our error bars on these quantities are larger than they quote,
despite our vastly improved galaxy sample. This is easily
explained. First, Keeton et al.\ (1998) used a fixed ($z_f = 15$)
evolutionary model and thus only fit for two parameters. The
significant correlations related to the evolution rate imply that
constraints on individual parameters will be much weaker when the
evolution rate is fit simultaneously. Second, Keeton et al.\ (1998)
individually converted each observed magnitude into a rest frame
magnitude, and used all values in the fit. Therefore, a galaxy was
represented by a number of data points equal to the number of bands in
which it was observed. By over-estimating the degrees of freedom, the
uncertainties were under-estimated.

\section{Constraints on the Star-formation Redshift}

We now constrain the mean star-formation redshift $\langle z_f
\rangle$ from our sample of 28 gravitational lens galaxies. This may
be accomplished in one of two ways. First, we can approximate the
predicted evolutionary tracks (Fig.~1) with a linear model ($d \log
(M/L) / dz \sim {\rm constant}$), and map the results of \S 3 and \S 4
into constraints on $\langle z_f \rangle$. Second, we can fit the lens
data to the predicted evolutionary tracks directly. Because the linear
evolution model becomes a poor approximation for $z_f < 1.5$, we
choose the latter. Note that the derived mean star-formation redshift
depends sensitively on the assumed value of the IMF slope (e.g., van
Dokkum \& Franx 1996), with steeper IMFs leading to lower $\langle z_f
\rangle$. The IMF is still quite uncertain (e.g., Scalo 1998), so we
simply fit to the evolutionary tracks of a Salpeter (1955) IMF to
facilitate comparison with other studies. As demonstrated by Treu et
al.\ (2001), changing to a Scalo (1986) IMF has little effect on the
derived range for the mean star-formation redshift.

The fits are performed using modified versions of Eq.~5 and 11, in
which the linear evolution model is replaced with the detailed
evolutionary tracks shown in Fig.~1. Constraints on the typical
star-formation redshift are obtained by evaluating $\chi^2$ as a
function of $\langle z_f \rangle$, assuming solar metallicity and an
instantaneous starburst. Uncertainties are handled as described in \S
3 and \S 4, and the $\chi^2$ is renormalized as before. Only $B$-band
results are described in detail.

Two trials are performed for the FP evolution. First, we calculate
mass-to-light ratio offsets using a fixed local FP intercept
(Fig.~8). The best-fit mean star-formation redshift is $\langle z_f
\rangle = 2.1$, and the favored $1\sigma$ ($2\sigma$) range, derived
from the $\chi^2$ analysis, is $1.9 < \langle z_f \rangle < 2.3$ ($1.8
< \langle z_f \rangle < 2.9$). Second, we allow the local FP intercept
to vary, which yields a best-fit of $\langle z_f \rangle = 2.3$. The
increased redshift is consistent with the slightly slower evolution
rates found in \S 3 when the intercept was re-fit. The $1\sigma$
($2\sigma$) permitted range is $2.0 < \langle z_f \rangle < 3.6$ ($
\langle z_f \rangle > 1.8$). Finally, a similar analysis using the FJ
relation yields a $1\sigma$ ($2\sigma$) bound of $\langle z_f \rangle
> 2.6$ ($\langle z_f \rangle > 1.7$). Constraints on $\langle z_f
\rangle$ are plotted in Fig.~9. Because the sample of 28 early-type
lens galaxies is consistent with old stellar populations, we have
little need to invoke significant episodes of secondary star-formation
at $z < 1$, as suggested by Treu et al.\ (2001, 2002).

\section{Discussion}

Gravitational lenses are the only sample of early-type galaxies
selected on the basis of mass, rather than luminosity-related
properties such as color, magnitude or surface brightness. The lens
sample should therefore be much less susceptible to Malmquist bias
than luminosity-selected samples, which will tend to over-represent
the brightest, bluest and youngest galaxies. Because such selection
problems grow more severe at higher redshifts, multi-color imaging
surveys are not well suited to extending galaxy samples beyond $z\sim
0.5$, where evolution effects are most pronounced. In contrast,
gravitational lensing naturally selects a galaxy sample out to $z\sim
1$, providing a long lever-arm with which to trace evolution. Lenses
are therefore excellent tools for investigating the evolution of
stellar populations in early-type field galaxies.

We have constrained the rate of luminosity evolution in a sample of 28
early-type gravitational lens galaxies spanning the redshift range $0
\la z \la 1$. These galaxies are estimated to have a mean stellar
velocity dispersion of $226\kms$ inside the Coma aperture, with a
standard deviation of $55\kms$. We reiterate that all evolution
results apply specifically to this mass scale.  First we investigated
the fundamental plane, comparing the inferred FP intercept for each
galaxy with that of the present-day $B$-band FP (Bender et al.\
1998). The lens sample implies an evolution rate of $d\log (M/L)_B/dz
= -0.56 \pm 0.04$ ($1 \sigma$) in the rest frame $B$ band. By
simultaneously fitting both the evolution rate and the present-day FP
intercept, we can minimize possible systematic effects related to the
assumed FP slopes and our technique for estimating stellar velocity
dispersions.  These fits yield a slightly slower evolution rate of $d
\log (M/L)_B / dz = - 0.54 \pm 0.09$, and a $z=0$ FP intercept of
$\gamma_0 = -8.88 - \log (h/0.5) \pm 0.04$. Finally, we investigated
evolution using the Faber-Jackson relation of gravitational lens
galaxies, which is closely related to corrected aperture mass-to-light
ratios. At rest frame $B$ band, we find a present-day characteristic
magnitude of $M_{*0} = -19.70 + 5\log h \pm 0.29$ (for $\sigma_{DM*} =
225 \kms$), a Faber-Jackson slope of $\gamma_{FJ} = 3.29 \pm 0.58$,
and an evolution rate of $d \log (M/L)_B/dz = -0.41 \pm 0.21$. The
evolution rates derived from each of the above measurements are
consistent, though the FJ relation offers a rather poor tracer of
evolution due to its large intrinsic scatter, and the necessity of
fitting three parameters.

The luminosity evolution of gravitational lenses suggests that the
stellar populations of early-type field galaxies formed at a mean
redshift of $\langle z_f \rangle > 1.8$ ($2\sigma$), assuming a
Salpeter IMF and a flat $\Omega_m = 0.3$ cosmology. Similar
conclusions are derived by van Dokkum et al.\ (2001), who measure an
evolution rate of $d \log (M/L)_B / dz = -0.59 \pm 0.15$ from a
luminosity-selected sample of field E/S0s. These FP evolution results
are consistent with a number of recent age estimates (Bernardi et al.\
1998; Schade et al.\ 1999; Im et al.\ 2002) which favor old,
passively-evolving stellar populations for early-type galaxies in the
field, thereby contradicting semi-analytic galaxy formation models
(e.g., Kauffmann 1996; Kauffmann \& Charlot 1998). Alternate
conclusions have been drawn by Treu et al.\ (2002), who measure an
evolution rate of $d\log (M/L)_B/dz = -0.72^{+0.11}_{-0.16}$ among
field E/S0s and favor $\langle z_f \rangle \sim 1$ (for a Salpeter or
Scalo IMF). We have re-analyzed the Treu et al.\ (2002) sample, in
part to check our own software, and confirm their claim of rapid
luminosity evolution. Interestingly, we derive an even faster
evolution rate from their sample if the local FP intercept is re-fit,
rather than fixed. To explain their rapid evolution, Treu et al.\
(2001, 2002) suggest that the luminosities of early-type field
galaxies may be boosted by secondary episodes of star formation at $z
< 1$. The claim is supported by the detection of significant [OII]
emission in 22\% of their sample, similar to the findings of Schade et
al.\ (1999) for field ellipticals out to $z\sim 1$. Menanteau, Abraham
\& Ellis (2001) further argue that internal color variations detected
in this galaxy population are consistent with recent star formation.
However, evidence for some recently-formed stellar populations does
not necessarily imply rapid luminosity evolution. For example, Schade
et al.\ (1999) detect copious [OII] emission, but still measure a slow
rate of luminosity evolution out to $z\sim 1$. [OII] emission of
various line-width has also been reported in several early-type
gravitational lens galaxies (e.g., Myers et al.\ 1995; Warren et al.\
1996; Fassnacht \& Cohen 1998; Tonry \& Kochanek 2000; Koopmans \&
Treu 2002a), even though the evolution rate of the lens sample implies
a rather high mean star-formation redshift.  Consequently, the
presence of [OII] emission does not provide a complete explanation for
the rapid rate of luminosity evolution derived by Treu et al.\
(2002). Additional data and analysis will therefore be necessary to
determine whether there is a discrepancy among the various samples of
intermediate-redshift early-type field galaxies, and if so, to
pinpoint its source.

Our estimates for the luminosity evolution rate and mean
star-formation epoch of gravitational lens galaxies have neglected the
role of progenitor bias (e.g., van Dokkum \& Franx 2001).\footnote{
Note that the field galaxy evolution rates of Treu et al.\ (2001,
2002) and van Dokkum et al.\ (2001) were not corrected for progenitor
bias, so our comparison of mass-selected and luminosity-selected
samples should be robust.} Because early-type galaxies are believed to
form via mergers of star-forming disk galaxies, the high-redshift E/S0
sample comprises only a subset of the progenitors of the local
sample. To properly constrain the evolution of stellar populations,
one should consider all progenitors of present-day early-type
galaxies. However, while the above analyses have considered only
early-type lenses, we expect that the inclusion of the rare late-type
lenses ($\sim 5/70$) would have little effect on the evolution rate
derived from our sample. Therefore, the absence of a significant
population of smaller separation, blue, star-forming late-type lens
galaxies argues against a strong progenitor bias in our estimation of
the evolution rate and star-formation redshift for mass-selected
early-type galaxies.

While gravitational lens galaxies are selected on the basis of mass,
those entering our analysis sample did need to satisfy a ``secondary
selection'' based on luminosity-related properties: the lens galaxies
must be bright enough (at least relative to the quasar components) to
obtain robust photometry. The effect of the secondary selection is
difficult to quantify, but like any luminosity-related selection, it
should lead to the preferential inclusion of brighter galaxies at
higher redshifts. Residual Malmquist bias may therefore result in a
slightly over-estimated evolution rate in our sample. Note, however,
that our ability to estimate stellar velocity dispersions should help
minimize the role of Malmquist bias. Velocity dispersions are much
more likely to be spectroscopically measured for brighter galaxies,
due to signal-to-noise constraints. Considering only those lens
galaxies with measured (or measurable) velocity dispersions would
therefore likely bias the sample toward faster evolution rates. The
first two lenses to be analyzed by the LSD survey (MG2016+112 and
0047--2808; KT) hint at this problem, as each of their implied
evolution rates is faster than that of the overall lens sample.
Still, the LSD survey will make a significant contribution to
evolution studies, particularly via its precision tests of the
isothermal hypothesis.

There is much work to be done in order to obtain a complete
evolutionary picture of the early-type galaxy population. One obvious
improvement would be larger samples of field E/S0s out to intermediate
redshift. Unfortunately, the early-type galaxy sample from the Sloan
Digital Sky Survey (Bernardi et al.\ 2001) is rather shallow ($z <
0.3$). The SDSS Luminous Red Galaxy Survey (e.g., Eisenstein et al.\
2001) aims to select a sample out to $z \sim 0.5$, which should be
more appropriate for evolution studies. This will be complemented by
the DEEP2 (Davis et al.\ 2001) galaxy samples at $z >
0.7$. Gravitational lenses will continue to play an important role due
to their unique status as the only mass-selected galaxies, and newly
discovered lenses provide raw material for expanding the current
sample. The primary obstacle will be obtaining the necessary HST
photometry and ground-based spectroscopy to turn these lenses into
viable astrophysical tools. Recent progress on these fronts has,
unfortunately, not been commensurate with the discovery rate. 

Vastly improved lens samples may one day allow us to pursue more
sophisticated analyses of luminosity evolution, such as its dependence
on velocity dispersion or color. Studies of environmental dependence
will be much more limited, however, as there are very few lenses like
Q0957+561, which reside in higher-density environments. Hence, we are
unlikely to produce a sample of mass-selected cluster galaxies
comparable to our current field sample. More promising are the clues
that gravitational lensing may offer regarding the morphological
history of field galaxies, for which present knowledge is still rather
limited (e.g., Im et al.\ 1999, 2002). If some high velocity
dispersion E/S0s are formed from the mergers of small velocity
dispersion disk galaxies at $z < 1$, this could potentially be traced
directly by the fraction of late-type lenses at different redshifts,
or indirectly by the redshift dependence of the mean image separation
(Mao \& Kochanek 1994; Rix et al.\ 1994). The current data do not
suggest any obvious trends, but the number of spiral lenses is small
because they are very inefficient deflectors (e.g., Kochanek
1996). Significantly larger samples of lenses will therefore be needed
to fully investigate the relationship between the luminosity and
morphological evolution of early-type galaxies in the field.

\acknowledgements 

We thank L. Koopmans, M. Pahre and T. Treu for useful conversations
regarding galaxy evolution; C. Fassnacht for his expertise on lens
galaxy spectra; and the referee, D. Kelson, for suggesting
improvements to the manuscript. This work is based on observations
made with the NASA/ESA Hubble Space Telescope, obtained at the Space
Telescope Science Institute, which is operated by AURA, Inc., under
NASA contract NAS 5-26555. We acknowledge the support of HST grants
GO-7495, 7887, 8175, 8804, and 9133. We acknowledge the support of the
Smithsonian Institution. CSK is supported by NASA ATP Grant NAG5-9265.

\appendix
\section{The Faber-Jackson Relation in Observed Bands}

Galaxy emission complicates gravitational lens surveys conducted at
optical or near-infrared wavelengths (e.g., Kochanek 1993a). To
properly interpret observed lensing rates, one must understand the
effect of galaxies on both the selection function and completeness. A
recent example of this involves calculations of the lensing rate among
color-selected high-redshift quasars in the Sloan Digital Sky Survey
(Wyithe \& Loeb 2002). If a foreground galaxy is multiply-imaging a
background quasar, the galaxy emission will affect the derived
colors. Hence, a bright lens galaxy can prevent a high-redshift lensed
quasar from meeting the color selection criteria. Such biases must be
removed by estimating the flux that a lensing galaxy will contribute
to a gravitational lens system of a given image separation. This is
best accomplished by applying the optical properties of a known sample
of gravitational lens galaxies.

We now provide a simple database for estimating the magnitude of an
early-type lensing galaxy in a wide range of filters. This is done by
determining the correlation among the redshift, the reduced image
separation, and the {\em effective absolute magnitude of lensing
galaxies in observed bands}: $M_{obs} = m_{obs} - DM$, where $DM$ is
the distance modulus. The combined effects of evolution and spectral
K-corrections are modeled using a single parameter $\gamma_{E+K}$,
such that the predicted magnitude is
\begin{equation}
M_{mod} = M_{*0} + 2.5 \gamma_{E+K} z_d - 1.25 \gamma_{FJ} \log \Delta
\theta_{red} \ .
\end{equation}
The fitting function is identical to Eq.~11. If the lens galaxy has
not been observed in a given filter, its magnitude is estimated from
the available filters using the spectrophotometric model (\S 2.2). If
data are available in that filter, the observed magnitude is used
directly. The fit parameters are presented in Table~3 for six HST
filters. Despite our crude parameterization of redshift effects, the
rms scatters in the relations are, on average, only slightly larger
than those in the rest frame bands.

\clearpage
\begin{deluxetable}{rrccccccc}
\tablecaption{ Lens Galaxy Photometric Data }
\tablewidth{0pt}
\tablehead{Lens & $z_d$ & $z_s$ & $E_{Gal}$ & $\Delta\theta$ & $\log \Delta\theta_{red}$ &{$\log(r_e/'')$} & Filter & Mag\\
 & & & (mag) & ($''$) & & & \\
}
\startdata
    0047--2808  & 0.49 & 3.60  & 0.016 & 2.70 & $+0.101$ &$-0.04\pm 0.04$ & F555W & $20.65\pm0.05$ \\
	& & & & & & & F814W & $18.92\pm0.27$\\

Q0142--100	& 0.49 & 2.72 & 0.032 & 2.24 & $+0.045$ & $-0.29 \pm 0.02$ & F160W & $16.63 \pm 0.03$ \\
 	& & & & & & & F814W & $18.72 \pm 0.05$\\
 	& & & & & & & F675W & $19.35 \pm 0.01$\\
 	& & & & & & & F555W & $20.81 \pm 0.02$\\

  CTQ414       & 0.28 & 1.29 &  0.015 &  1.22 & $-0.238$ & $-0.18\pm0.06$ & F160W & $16.67\pm0.15$ \\
	& $\pm0.02$ & & & & $\pm 0.017$ & & F814W & $18.91\pm0.37$\\  
	& & & & & & & F555W & $20.36\pm0.18$\\

   MG0414+0534  & 0.96 & 2.64  & 0.303 & 2.38 & $+0.246$ &$-0.11\pm 0.08$ & F160W & $17.54\pm0.14$ \\
	& & & & & & & F205W & $16.70\pm0.12$\\  			      
	& & & & & & & F110W & $19.21\pm0.03$\\
	& & & & & & & F814W & $20.91\pm0.05$\\ 				      
	& & & & & & & F675W & $22.58\pm0.13$\\  			      
	& & & & & & & F555W & $24.17\pm0.15$\\				      

     B0712+472  & 0.41 & 1.34  & 0.113 & 1.42 & $-0.096$ &$-0.44\pm 0.06$ & F160W & $17.16\pm0.15$ \\ 
	& & & & & & & F814W & $19.56\pm0.07$\\
	& & & & & & & F555W & $21.75\pm0.10$\\

   HS0818+1227  & 0.39 & 3.12  & 0.031 & 2.83 & $+0.102$ &$-0.05\pm 0.01$ & F814W & $18.76\pm0.02$ \\
	& & & & & & & F555W & $20.78\pm0.10$\\

  FBQ0951+2635 & 0.20 & 1.24 &  0.022 &  1.11 & $-0.315$ & $-0.78\pm0.11$ & F160W & $17.86\pm0.23$ \\
	& $\pm0.02$ & & & & $\pm 0.012$ & & F814W & $19.67\pm0.23$\\ 
	& & & & & & & F555W & $21.02\pm0.20$\\

 BRI0952--0115 & 0.38 & 4.50 &  0.063 &  1.00 & $-0.371$ & $-1.00\pm0.12$ & F160W & $18.95\pm0.16$ \\
	& $\pm0.03$ & & & & $\pm 0.008$ & & F814W & $21.21\pm0.04$\\ 
	& & & & & & & F675W & $22.08\pm0.03$\\ 
	& & & & & & & F555W & $23.67\pm0.08$\\
LBQS1009--0252	& 0.78 & 2.74  & 0.034 & 1.54 & $-0.039$ & $-0.75 \pm 0.06$ & F160W & $19.30 \pm 0.12$ \\
 	& $\pm0.10$& & & & $\pm 0.032$ & & F814W & $21.99 \pm 0.04$\\
	& & & & & & & F555W & $24.05 \pm 0.25$\\

\hline
\tablebreak

    Q1017--207 & 0.86 & 2.55 &  0.046 &  0.85 & $-0.217$ & $-0.52\pm0.01$ & F160W & $19.26\pm0.06$ \\
	& $\pm0.15$ & & & & $\pm 0.048$ & & F814W & $21.82\pm0.48$\\ 
	& & & & & & & F555W & $25.48\pm0.73$\\

 FSC10214+4724 & 0.96 & 2.29 &  0.012 &  1.59 & $+0.091$ & $+0.06\pm0.19$ & F814W & $20.40\pm0.39$ \\
	& $\pm0.10$ & & & & $\pm0.044$ & & F205W & $17.04\pm0.40$\\ 
	& & & & & & & F110W & $19.29\pm0.37$\\ 
	& & & & & & & F555W & $23.18\pm0.62$\\

     B1030+074  & 0.60 & 1.54  & 0.022 & 1.56 & $+0.028$ &$-0.35\pm 0.06$ & F814W & $20.24\pm0.13$ \\
	& & & & & & & F160W & $17.64\pm0.15$\\				      
	& & & & & & & F555W & $22.71\pm0.12$\\ 				      

  HE1104--1805  & 0.73 & 2.32  & 0.056 & 3.19 & $+0.309$ &$-0.20\pm 0.13$ & F160W & $17.47\pm0.27$ \\
	& & & & & & & F814W & $20.01\pm0.10$\\ 				      
	& & & & & & & F555W & $23.14\pm0.58$\\				      

    PG1115+080  & 0.31 & 1.72  & 0.041 & 2.29 & $+0.026$ &$-0.33\pm 0.02$ & F160W & $16.66\pm0.04$ \\
	& & & & & & & F814W & $18.92\pm0.02$\\
	& & & & & & & F555W & $20.74\pm0.04$\\

   MG1131+0456  & 0.84 & -- &   0.036 &  2.10 & $+0.214$ & $-0.24\pm0.05$ & F160W & $18.62\pm0.08$ \\   
	& & & & & $\pm 0.127$ & & F814W & $21.21\pm0.04$\\ 		  
	& & & & & & & F675W & $22.47\pm0.06$\\  		  
	& & & & & & & F555W & $23.85\pm0.24$\\		  

 HST14113+5211  & 0.46 & 2.81  & 0.016 & 1.72 & $-0.083$ &$-0.33\pm 0.05$ & F702W & $20.50\pm0.06$ \\
	& & & & & & & F814W & $19.99\pm0.03$\\
	& & & & & & & F555W & $22.19\pm0.06$\\
						
 HST14176+5226  & 0.81 & 3.40  & 0.007 & 2.84 & $+0.225$ &$-0.15\pm 0.05$ & F160W & $17.53\pm0.11$ \\
	& & & & & & & F814W & $19.77\pm0.06$\\				      
	& & & & & & & F606W & $21.91\pm0.06$\\				      
     B1422+231  & 0.34 & 3.62  & 0.048 & 1.56 & $-0.177$ &$-0.50\pm 0.13$ & F160W & $17.57\pm0.20$ \\
	& & & & & & & F791W & $19.66\pm0.25$\\
	& & & & & & & F555W & $21.80\pm0.17$\\

\hline
\tablebreak
\tablebreak

   SBS1520+530  & 0.72 & 1.86  & 0.016 & 1.59 & $+0.047$ &$-0.46\pm 0.04$ & F160W & $17.84\pm0.06$ \\
	& & & & & & & F814W & $20.16\pm0.13$\\				      
	& & & & & & & F555W & $21.96\pm1.24$\\ 				      
									      
   MG1549+3047  & 0.11 & -- &   0.029 &  2.30 & $-0.061$ & $-0.09\pm0.04$ & F160W & $14.73\pm0.07$ \\
	& & & & & $\pm 0.009$ & & F205W & $14.00\pm0.01$\\  
	& & & & & & & F814W & $16.70\pm0.02$\\ 
	& & & & & & & F555W & $18.19\pm0.02$\\  

     B1608+656  & 0.63 & 1.39  & 0.031 & 2.27 & $+0.242$ &$-0.19\pm 0.07$ & F160W & $16.76\pm0.13$ \\
	& & & & & & & F814W & $19.02\pm0.13$\\ 				      
	& & & & & & & F555W & $21.24\pm0.23$\\				      

   MG1654+1346  & 0.25 & 1.74  & 0.061 & 2.10 & $-0.039$ &$-0.05\pm 0.01$ & F160W & $15.83\pm0.04$ \\
	& & & & & & & F814W & $17.90\pm0.02$\\
	& & & & & & & F675W & $18.55\pm0.01$\\
	& & & & & & & F555W & $19.72\pm0.03$\\

     B1938+666	& 0.88 &  --  & 0.121 & 1.00 & $-0.088$ & $-0.16 \pm 0.04$ & F160W & $18.67 \pm 0.08$ \\
 	& & & & & $\pm0.154$ & & F814W & $21.46 \pm 0.08$\\
 	& & & & & & & F555W & $24.45 \pm 0.84$\\

    MG2016+112  & 1.00 & 3.27  & 0.235 & 3.26 & $+0.354$ &$-0.66\pm 0.05$ & F160W & $18.46\pm0.09$ \\
	& & & & & & & F814W & $21.95\pm0.09$\\				      
	& & & & & & & F555W & $25.12\pm1.06$\\				      

     B2045+265  & 0.87 & 1.28$^{a}$ & 0.232 &  2.28 & $+0.265$ & $-0.42\pm0.14$ & F160W & $18.25\pm0.26$ \\  
	& & & & & $\pm 0.145$ & & F814W & $21.25\pm0.19$\\  		  
	& & & & & & & F555W & $23.86\pm0.22$\\		  

  HE2149--2745  & 0.50 & 2.03  & 0.032 & 1.70 & $-0.039$ &$-0.30\pm 0.04$ & F160W & $17.61\pm0.10$ \\
	& & & & & & & F814W & $19.56\pm0.03$\\
	& & & & & & & F555W & $21.18\pm0.09$\\
\hline
\tablebreak
\tablebreak
\tablebreak
\tablebreak
\tablebreak
     Q2237+030  & 0.04 & 1.69  & 0.071 & 1.76 & $-0.202$ &$+0.59\pm 0.08$ & F160W & $12.22\pm0.22$ \\
	& & & & & & & F205W & $11.88\pm0.13$\\
	& & & & & & & F814W & $14.15\pm0.20$\\
	& & & & & & & F675W & $14.66\pm0.23$\\ 
	& & & & & & & F555W & $15.49\pm0.22$\\   

     B2319+051  & 0.62 & -- &   0.064 &  1.36 & $-0.080$ & $-0.65\pm0.01$ & F160W & $18.20\pm0.02$ \\
	& & & & & $\pm 0.076$ & & F814W & $20.71\pm0.77$\\  		  
	& & & & & & & F555W & $23.43\pm0.04$\\  		  

\enddata \tablecomments{Listed for each lens are the galaxy ($z_d$)
and source ($z_s$) redshifts, Galactic extinction ($E_{Gal} \equiv
E(B-V)$) from Schlegel et al.\ (1998), standard image separation
($\Delta\theta$), reduced image separation ($\log \Delta
\theta_{red}$), intermediate axis effective radius ($\log (r_e/'')$),
and magnitudes in each filter. The first filter listed has the highest
signal-to-noise ratio, and the surface brightness ($\mu_e$) and
effective radius are simultaneously fit using a de Vaucouleurs
profile. The effective radius is then held fixed when determining
$\mu_e$ in the remaining filters. The listed magnitudes are
extrapolated from the fit quantities, $m = \mu_e - 5 \log r_e - 2.5
\log 2\pi$, and their errors reflect the fitting technique used in
each filter. Magnitudes have {\em not} been corrected for Galactic
extinction in this table, but an $R_V = 3.1$ extinction curve is used
to correct the magnitudes in all calculations and figures. The values
of $\log \Delta \theta_{red}$ are calculated assuming $\Omega_M =
0.3$, $\Omega_{\Lambda} = 0.7$. Galaxy redshift errors indicate lenses
on which the FP redshift estimation technique has been applied (\S
2.4). For lenses with unmeasured galaxy or source redshifts, errors on
$\log \Delta \theta_{red}$ are constructed using Monte Carlo methods
(\S 2.5).  If $z_d$ is estimated, correlation between $z_d$ and $\log
\Delta \theta_{red}$ is important, but here we list the errors
obtained by assuming uncorrelated quantities. $^{a}$ The B2045+265
(Fassnacht et al.\ 1999) source redshift is currently disputed, as it
implies an uncomfortably large velocity dispersion for the lens
galaxy. We therefore treat it as an unknown, and set $z_s = 2.0 \pm
1.0$ as we have done for lenses with unmeasured source redshifts.}
\end{deluxetable}

\clearpage
\begin{deluxetable}{lcccc}
\tablecolumns{5}
\tablecaption{Evolution Estimates from the Faber-Jackson Relation}
\tablehead{ \colhead{Rest Band} &\colhead{$M_{*0} - 5 \log h$} & \colhead{$\gamma_{FJ}$} & \colhead{$\gamma_{EV}$} & rms}
\startdata 
$B$   		& $-19.70 \pm 0.29$ & $3.29 \pm 0.58$  & $-0.41\pm 0.21$ & 0.53 \\
F555W = $V$     & $-20.67 \pm 0.29$ & $3.29 \pm 0.58$  & $-0.36\pm 0.21$ & 0.53 \\
F814W = $I$  	& $-21.86 \pm 0.29$ & $3.30 \pm 0.59$  & $-0.31\pm 0.21$ & 0.53 \\
F160W = $H$   	& $-23.74 \pm 0.29$ & $3.34 \pm 0.60$  & $-0.21\pm 0.22$ & 0.53 \\
\enddata
\tablecomments{Listed for each rest frame band are the characteristic
magnitude at $z=0$ ($M_{*0} - 5 \log h$), FJ slope ($\gamma_{FJ}$),
evolution rate ($\gamma_{EV} \equiv d\log(M/L)/dz$), and unweighted
rms scatter about the best-fit relation.}
\end{deluxetable}

\begin{deluxetable}{lcccc}
\tablecolumns{5}
\tablecaption{Predicting Observed Lens Galaxy Magnitudes from the Faber-Jackson Relation}
\tablehead{ \colhead{Obs Band} &\colhead{$M_{*0} - 5 \log h$} & \colhead{$\gamma_{FJ}$} & \colhead{$\gamma_{E+K}$} & rms}
\startdata 
F555W = $V$   		& $-20.75 \pm 0.37$ & $3.13 \pm 0.83$  & $+0.84\pm 0.28$ & 0.63\\
F675W = $R$  		& $-21.86 \pm 0.30$ & $3.43 \pm 0.65$  & $+0.73\pm 0.22$ & 0.56\\
F814W = $I$  		& $-21.98 \pm 0.30$ & $3.21 \pm 0.65$  & $+0.16\pm 0.23$ & 0.54\\
F110W = $J$  		& $-22.75 \pm 0.28$ & $3.32 \pm 0.57$  & $-0.12\pm 0.20$ & 0.51\\
F160W = $H$  		& $-23.79 \pm 0.28$ & $3.39 \pm 0.57$  & $-0.21\pm 0.21$ & 0.52\\
F205W = $K$  		& $-24.26 \pm 0.27$ & $3.32 \pm 0.57$  & $-0.39\pm 0.20$ & 0.52\\
\enddata
\tablecomments{Listed for each observed band are the characteristic
magnitude at $z=0$ ($M_{*0} - 5 \log h$), FJ slope ($\gamma_{FJ}$),
combined evolution plus K-correction parameter ($\gamma_{E+K}$), and
unweighted rms scatter about the best-fit relation.}
\end{deluxetable}

\clearpage

\begin{figure*}
\begin{tabular}{c}
\psfig{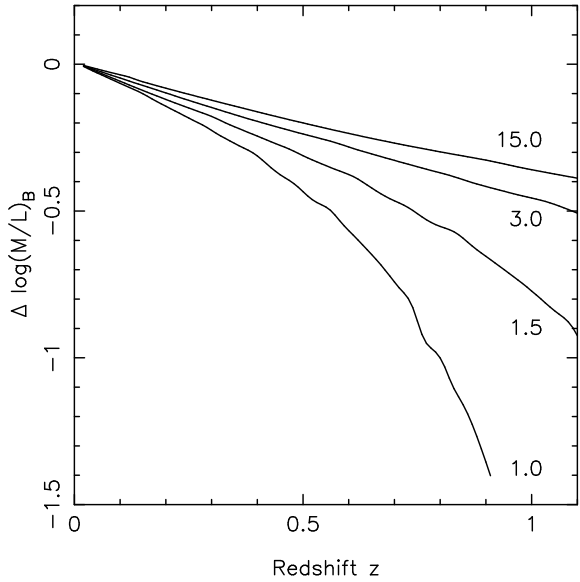}\\
\end{tabular}
\figurenum{1}
\caption{Pure luminosity evolution at rest frame $B$ band. Plotted is
the change in $\log (M/L)_B$ from its present-day value. The stellar
populations have solar metallicity and were formed in an instantaneous
burst. A Salpeter IMF is assumed. Tracks are computed with the
GISSEL96 version of the Bruzual \& Charlot (1993) spectral evolution
models, assuming a flat $\Omega_m = 0.3$ cosmology and $h = 0.65$. The
predicted evolutionary tracks for star-formation redshifts of $z_f =
15.0$, $3.0$, $1.5$ and $1.0$ are displayed. Note that later
star-formation redshifts result in faster evolution rates. For $z_f >
1.5$, the evolution rate at $z<1$ can be parameterized as linear in
redshift.}
\end{figure*}

\begin{figure*}
\begin{tabular}{c}
\psfig{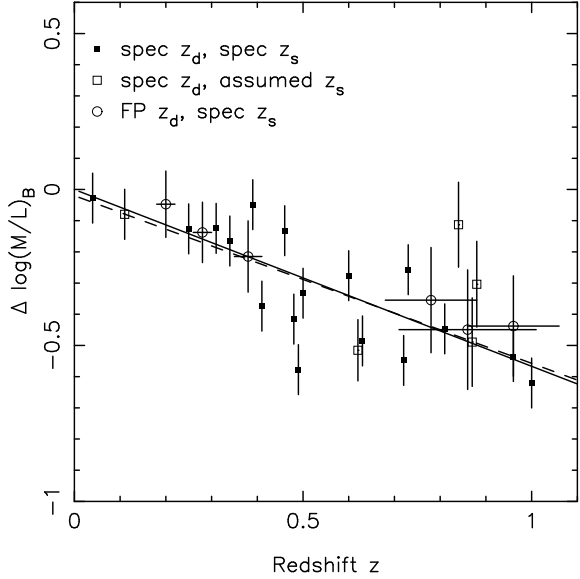}\\
\end{tabular}
\figurenum{2}
\caption{Evolution of lens galaxies from the FP at rest frame $B$
band. Plotted is the offset in $\log (M/L)_B$ derived using the offset
from the local FP intercept of $\gamma_0 = -8.895 - \log (h/0.5) \pm
0.01$. FP slopes of $\alpha = 1.25$ and $\beta = 0.32$ are
assumed. Lenses with known galaxy ($z_d$) and source ($z_s$) redshifts
are plotted as filled squares. Lenses with known $z_d$ and assumed
$z_s=2.0 \pm 1.0$ are plotted as open squares. Lenses with estimated
$z_d$ and known $z_s$ are plotted as open circles. Vertical and
horizontal error bars are correlated. The solid line depicts the
best-fit slope of $d \log (M/L)_B / dz= -0.56 \pm 0.04$ when
$\gamma_0$ is fixed. The dashed line depicts the best-fit slope of $d
\log (M/L)_B / dz = -0.54 \pm 0.09$ when $\gamma_0$ is simultaneously
re-fit.}
\end{figure*}

\begin{figure*}
\begin{tabular}{c}
\psfig{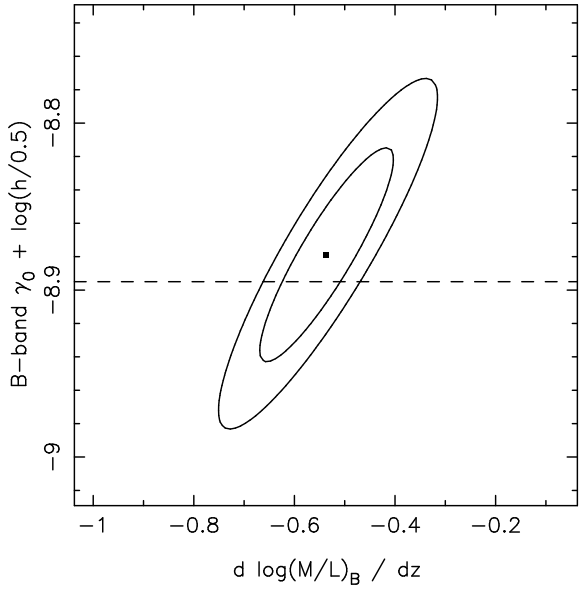}\\
\end{tabular}
\figurenum{3}
\caption{Simultaneous constraints on the evolution rate and
present-day FP intercept $\gamma_0$ at rest frame $B$ band. The FP
slopes are fixed at $\alpha = 1.25$ and $\beta = 0.32$.  The contours
enclose the $1 \sigma$ ($\Delta \chi^2 < 2.30$) and $2 \sigma$
($\Delta \chi^2 < 6.17$) confidence regions. The horizontal dashed
line represents the local (Coma cluster) FP intercept of $-8.895 -
\log(h/0.5)$.}
\end{figure*}

\begin{figure*}
\begin{tabular}{c}
\psfig{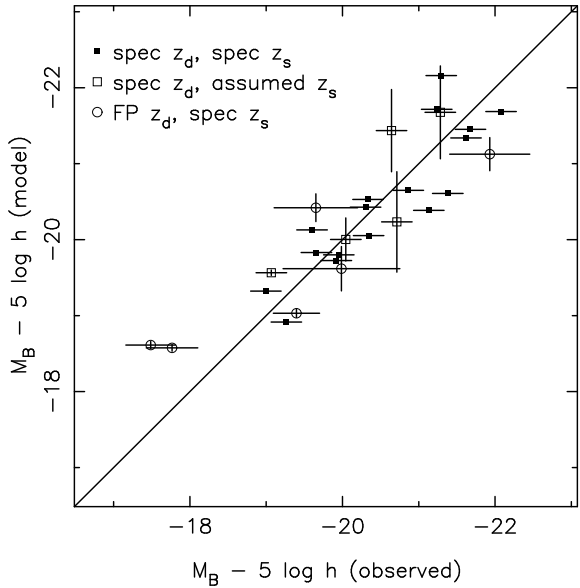}\\
\end{tabular}
\figurenum{4}
\caption{The Faber-Jackson relation for gravitational lenses. Plotted
is the $B$-band galaxy magnitude versus the prediction of the best-fit
model. Lenses with different redshift information are marked as in
Fig.~2.}
\end{figure*}

\begin{figure*}
\begin{tabular}{c}
\psfig{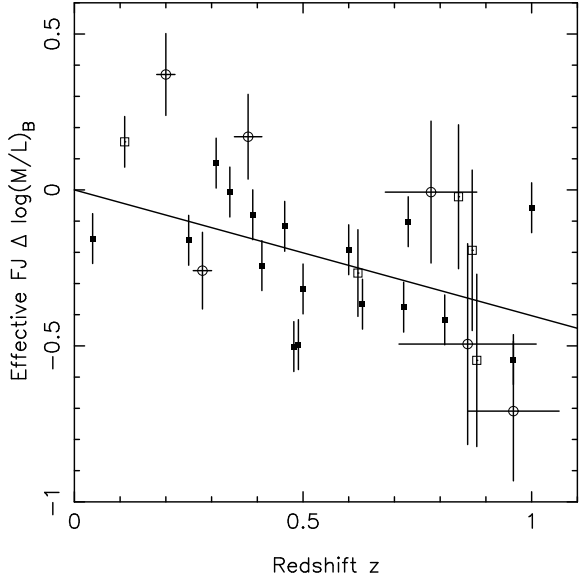}\\
\end{tabular}
\figurenum{5}
\caption{Evolution of the effective Faber-Jackson $B$-band
mass-to-light ratio. The scale is arbitrary. The best-fit slope is $d
\log (M/L)_B / dz = -0.41 \pm 0.21$. Lenses with different redshift
information are marked as in Fig.~2.}
\end{figure*}

\begin{figure*}
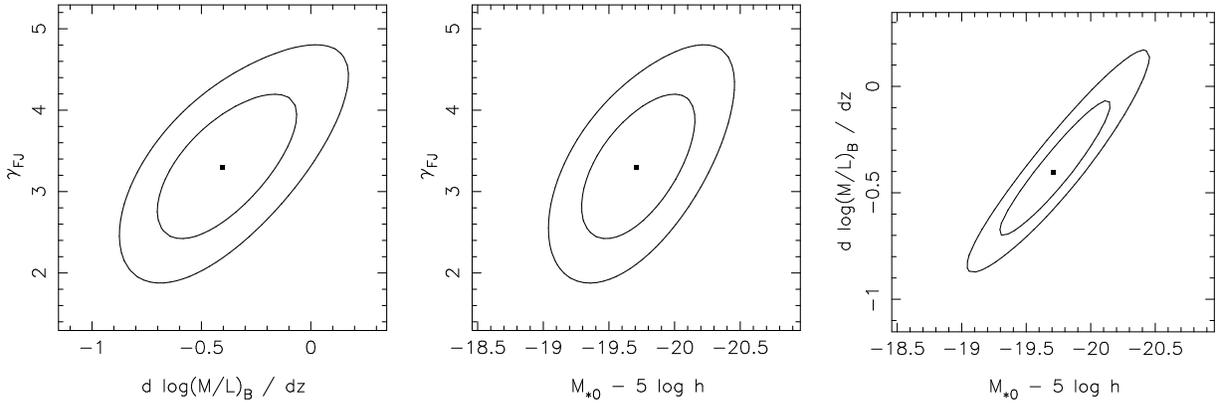

\begin{tabular}{ccc}
\psfig{file=newf6a.ps,width=2in}&
\psfig{file=newf6b.ps,width=2in}&
\psfig{file=newf6c.ps,width=2in}\\
\end{tabular}
\figurenum{6}
\caption{Pairwise constraints for the $B$-band values of $M_{*0} - 5
\log h$, $\gamma_{FJ}$, and $d\log(M/L)_B/dz$ from the Faber-Jackson
relation. The contours enclose the $1 \sigma$ ($\Delta \chi^2 < 2.30$)
and $2 \sigma$ ($\Delta \chi^2 < 6.17$) confidence regions.}
\end{figure*}

\begin{figure*}
\begin{tabular}{c}
\psfig{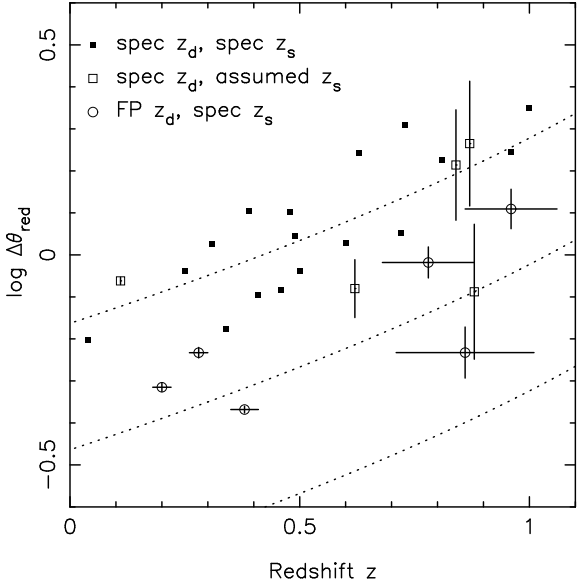}\\
\end{tabular}
\figurenum{7}
\caption{The reduced image separation $\log \Delta \theta_{red}$ as a
function of redshift. Dotted lines correspond to constant angular
image separations for a source at $z_s = 2.0$ (bottom to top: $\Delta
\theta = 0\farcs5$, $1\farcs0$, $2\farcs0$). Lenses with different
redshift information are marked as in Fig.~2. The absence of low
redshift lens galaxies with large $\log \Delta \theta_{red}$ is
puzzling.  }
\end{figure*}

\begin{figure*}
\begin{tabular}{c}
\psfig{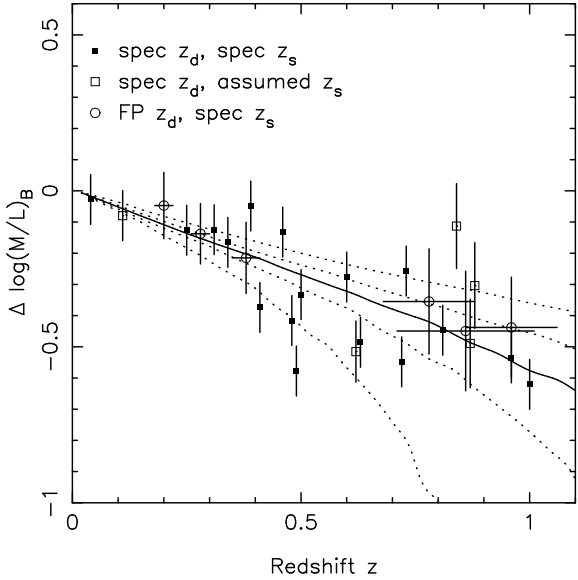}\\
\end{tabular}
\figurenum{8}
\caption{Direct fits to evolutionary tracks for a Salpeter
IMF. Results are shown for $B$-band FP mass-to-light ratios calculated
using a fixed local intercept. Therefore, the plotted data points are
identical to those in Fig.~2. An instantaneous starburst and solar
metallicity are assumed. Dotted lines represent star-formation
redshifts of $z_f = 15.0$, $3.0$, $1.5$ and $1.0$ (top to bottom). The
solid line represents the best-fit track of $\langle z_f \rangle =
2.1$.}
\end{figure*}

\begin{figure*}
\begin{tabular}{c}
\psfig{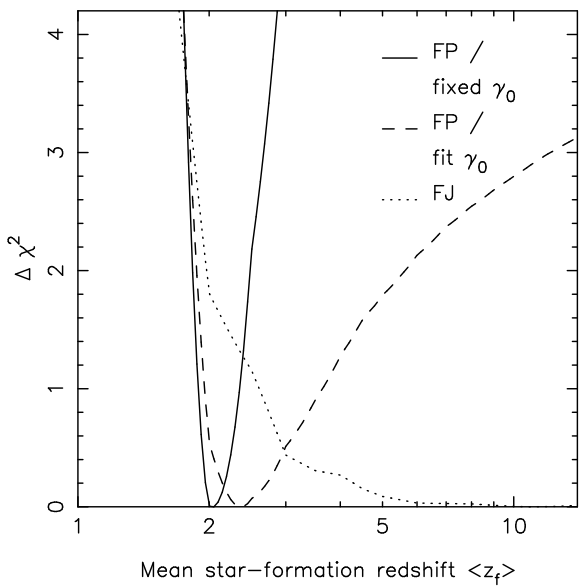}\\
\end{tabular}
\figurenum{9}
\caption{Constraints on the mean star-formation redshift $\langle z_f
\rangle$ for a Salpeter IMF.  Plotted is the renormalized $\Delta
\chi^2$ as a function of $\langle z_f \rangle$. The solid (dashed)
line is derived from the FP analysis in which the $z=0$ intercept
$\gamma_0$ is held fixed (simultaneously re-fit). The dotted line is
derived from the FJ analysis. A mean star-formation redshift of
$\langle z_f \rangle > 1.8$ is favored at $2 \sigma$ confidence.}
\end{figure*}


\clearpage
\begin{references}

\reference{}
Bade, N., Siebert, J., Lopez, S., Voges, W., \& Reimers, D. 1997,
A\&A, 317L, 13

\reference{Baugh}
Baugh, C.M., Cole, S., \& Frenk, C.S. 1996, MNRAS, 283, 1361

\reference{Bender $B$ band} 
Bender, R., Saglia, R.P., Ziegler, B., Belloni, P., Greggio, L., Hopp,
U., \& Bruzual, G. 1998, ApJ, 493, 529

\reference{}
Bernardi, M., Renzini, A., da Costa, L.N., Wegner, G., Alonso, M.V.,  
Pellegrini, P.S., Rit\'e, C., \& Willmer, C.N.A. 1998, ApJL, 508, L143

\reference{Bernardi}
Bernardi, M., et al.\ 2001, AJ, submitted (astro-ph/0110344)

\reference{Binney} 
Binney, J., \& Tremaine, S. 1987, Galactic Dynamics (Princeton:
University Press)

\reference{}
Bower, R.G., Lucey, J.R., \& Ellis, R.S. 1992, MNRAS, 254, 601

\reference{stellar models}
Bruzual, A.G., \& Charlot, S. 1993, ApJ, 405, 538

\reference{}
Cardelli, J.A., Clayton, G.C., \& Mathis, J.S. 1989, ApJ, 345, 245

\reference{} 
Cohn, J.D., Kochanek, C.S., McLeod, B.A., \& Keeton, C.R. 2001, ApJ,
554, 1216

\reference{}
Cole, S., Aragon-Salamanca, A., Frenk, C.S., Navarro, J.F., \&  Zepf, S.E.
1994, MNRAS, 271, 781


\reference{cluster interactions 2} 
Couch, W.J., Barger, A.J., Smail, I., Ellis, R.S., \& Sharples,
R.M. 1998, ApJ, 497, 188

\reference{Clustering 2}
Davis, M., Efstathiou, G., Frenk, C.S., \& White, S.D.M. 1985, ApJ,
292, 371

\reference{}
Davis, M., Newman, J.A., Faber, S.M., \& Phillips, A.C. 2001, in
Proc.\ of the ESO/ECF/STScI Workshop on Deep Fields,
eds. S. Cristiani, A. Renzini \& R.E. Williams (Springer), 241

\reference{}
de Vaucouleurs, G. 1961, ApJS, 5, 233

\reference{FP1} 
Djorgovski, S.G., \& Davis, M. 1987, ApJ, 313, 59

\reference{FP2} 
Dressler, A., Lynden-Bell, D., Burstein, D., Davies, R.J., Faber, S.M.,
Terlevich, R.J., \& Wegner, G. 1987, ApJ, 313, 42

\reference{cluster interactions 3} 
Dressler, A., Oemler, A., Jr., Sparks, W.B., \& Lucas, R.A. 1994,
ApJL, 435, L23

\reference{}
Dressler, A., et al.\ 1997, ApJ, 490, 577

\reference{}
Efstathiou, G., Ellis, R.S., \& Peterson, B.A. 1988, MNRAS, 232, 431

\reference{}
Eisenstein, D.J., et al.\ 2001, AJ, 122, 2267

\reference{}
Ellis, R.S., Smail, I., Dressler, A., Couch, W.J.,
Oemler, A., Jr., Butcher, H., \&  Sharples, R.M. 1997, ApJ, 483, 582

\reference{}
Fabbiano, G. 1989, ARA\&A, 27, 87

\reference{}
Faber, S.M. 1973, ApJ, 179, 731

\reference{FJ} 
Faber, S.M., \& Jackson, R.E. 1976, ApJ, 204, 668

\reference{}
Fabricant, D., Franx, M., \& van Dokkum, P.G. 2000, ApJ, 539, 577


\reference{}
Fassnacht, C.D., \& Cohen, J.G. 1998, AJ, 115, 377

\reference{}
Fassnacht, C.D., et al.\ 1999, AJ, 117, 658

\reference{}
Ferreras, I., Charlot, S., \& Silk, J. 1999, ApJ, 521, 81

\reference{}
Foltz, C.B., Hewett, P.C., Webster, R.L., \& Lewis, G.F. 1992, ApJL,
386, L43

\reference{FJ}
Forbes, D.A., \& Ponman, T.J. 1999, MNRAS, 309, 623

\reference{}
Gerhard, O., Kronawitter, A., Saglia, R.P., \& Bender, R. 2001, AJ,
121, 1936

\reference{HS0818}
Hagen, H.-J., \& Reimers, D. 2000, A\&A, 357L, 29

\reference{Hernquist}
Hernquist, L. 1990, ApJ, 356, 359

\reference{}
Holtzman, J.A., Burrows, C.J., Casertano, S., Hester, J.J.; Trauger,
J.T., Watson, A.M., \& Worthey, G. 1995, PASP, 107, 1065

\reference{2237}
Huchra, J., Gorenstein, M., Kent, S., Shapiro, I., Smith, G., Horine,
E., \& Perley, R. 1985, AJ, 90, 691

\reference{}
Im, M., Griffiths, R.E., Naim, A., Ratnatunga, K.U., Roche, N., Green,
R.F., \& Sarajedini, V.L. 1999, ApJ, 510, 82

\reference{}
Im, M., et al.\ 2002, ApJ, 571, 136

\reference{Coma}
Jorgensen, I., Franx, M., \& Kjaergaard, P. 1996, MNRAS, 280, 167


\reference{FP evolution}
Jorgensen, I., Franx, M., Hjorth, J., \& van Dokkum, P.G. 1999, MNRAS,
308, 833

\reference{Clustering 4} 
Kauffmann, G., White, S.D.M., \& Guiderdoni, B. 1993, MNRAS, 264, 201


\reference{Clustering 3} 
Kauffmann, G. 1996, MNRAS, 281, 487

\reference{Kauffmann 2}
Kauffmann, G., \& Charlot, S. 1998, MNRAS, 294, 705

\reference{} 
Keeton, C.R., Kochanek, C.S., \& Falco, E.E. 1998, ApJ, 509, 561

\reference{}
Keeton, C.R., Christlein, D., \& Zabludoff, A.I. 2000, ApJ, 545, 129

\reference{}
Kelson, D.D., van Dokkum, P.G., Franx, M., Illingworth, G.D., \&
Fabricant, D. 1997, ApJL, 478, L13

\reference{}
Kelson, D.D., Illingworth, G.D., van Dokkum, P.G., \& Franx, M. 2000,
ApJ, 531, 184




\reference{}
Kochanek, C.S. 1993a, ApJ, 417, 438

\reference{}
Kochanek, C.S. 1993b, ApJ, 419, 12

\reference{}
Kochanek, C.S. 1994, ApJ, 436, 56

\reference{} 
Kochanek, C.S. 1995, ApJ, 445, 559

\reference{}
Kochanek, C.S. 1996, ApJ, 466, 638

\reference{} 
Kochanek, C.S., et al.\ 2000, ApJ, 543, 131

\reference{1127}
Koopmans, L.V.E., et al.\ 1999, MNRAS, 303, 727

\reference{2016 dispersion}
Koopmans, L.V.E., \& Treu, T. 2002a, ApJL, 568, L5

\reference{0047 KT}
Koopmans, L.V.E., \& Treu, T. 2000b, ApJ, in press (astro-ph/0205281)


\reference{cluster interactions 4} 
Lavery, R.J., \& Henry, J.P. 1988, ApJ, 330, 596

\reference{2016 discovery}
Lawrence, C.R., Schneider, D.P., Schmidet, M., Bennett, C.L.,
Hewitt, J.N., Burke, B.F., Turner, E.L., \& Gunn, J.E. 1984, Sci, 223, 46

\reference{}
Leh\'ar, J., Cooke, A.J., Lawrence, C.R., Silber, A.D., \& Langston,
G.I. 1996, AJ, 111, 1812

\reference{}
Leh\'ar, J., et al.\ 2000, ApJ, 536, 584


\reference{Lilly}
Lilly, S.J., Tresse, L., Hammer, F., Crampton, D., \& Le Fevre,
O. 1995, ApJ, 455, 108

\reference{CNOC}
Lin, H., Yee, H.K.C., Carlberg, R.G., Morris, S.L., Sawicki, M.,
Patton, D.R., Wirth, G., \& Shepherd, C.W. 1999, ApJ, 518, 533

\reference{}
Lubin, L.M., Fassnacht, C.D., Readhead, A.C.S., Blandford, R.D., \&
Kundic, T. 2000, AJ, 119, 451


\reference{}
Lucey, J.R., Bower, R.G., \& Ellis, R.S. 1991, MNRAS, 249, 755

\reference{}
Madau, P., Pozzetti, L., \& Dickinson, M. 1998, ApJ, 498, 106

\reference{}
Madgwick, D.S., et al.\ 2002, MNRAS, 333, 133

\reference{}
Mao, S.D., \& Kochanek, C.S. 1994, MNRAS, 268, 569

\reference{}
Menanteau, F., Abraham, R.G., \& Ellis, R.S. 2001, MNRAS, 322, 1

\reference{CTQ414}
Morgan, N.D., Dressler, A., Maza, J., Schechter, P.L., \&
Winn, J.N. 1999, AJ, 118, 1444

\reference{}
Myers, S.T. et al.\ 1995, ApJL, 447, L5

\reference{}
Mu\~noz, J.A., Kochanek, C.S., \& Keeton, C.R. 2001, ApJ, 558, 657


\reference{near IR plane}
Pahre, M.A., Djorgovski, S.G., \& de Carvalho, R.R. 1998a, AJ, 116, 1591

\reference{theory}
Pahre, M.A., de Carvalho, R.R., \& Djorgovski, S.G. 1998b, AJ, 116, 1606

\reference{Pahre: evolution in K alpha}
Pahre, M.A., Djorgovski, S.G., \& de Carvalho, R.R. 2001, Ap\&SS, 276, 983

\reference{}
Press, W.H., Teukolsky, S.A., Vetterling, W.T., \& Flannery,
B.P. 1992, Numerical Recipes (Cambridge: University Press)

\reference{HST12531-2914}
Ratnatunga, K.U., Ostrander, E.J., Griffiths, R.E., \& Im, M. 1995,
ApJL, 453, L5


\reference{}
Rix, H.-W., Maoz, D., Turner, E.L., \& Fukugita, M. 1994, ApJ, 435, 49

\reference{} 
Rix, H.-W., de Zeeuw, P.T., Cretton, N., van der Marel, R.P., \&
Carollo, C.M. 1997, ApJ, 488, 702


\reference{} 
Rusin, D., Norbury, M., Biggs, A.D., Marlow, D.R., Jackson, N.J.,
Browne, I.W.A., Wilkinson, P.N., \& Myers, S.T. 2002, MNRAS, 330, 205

\reference{}
Salpeter, E. 1955, ApJ, 121, 161

\reference{}
Scalo, J. 1986, Fund. Cosmic Phys., 11, 1

\reference{} 
Scalo, J. 1998, in ASP Conf. Ser. 142, The Stellar Initial Mass
Function: 38th Herstmonceux Conf., ed. G. Gilmore \& D. Howell (San
Francisco: ASP), 201

\reference{}
Schade, D., et al.\ 1999, ApJ, 525, 31

\reference{dust}
Schlegel, D.J., Finkbeiner, D.P., \& Davis, M. 1998, ApJ, 500, 525


\reference{SEF} 
Schneider, P., Ehlers, J., \& Falco, E.E. 1992, Gravitational Lenses
(Berlin: Springer-Verlag)

\reference{beta no change} 
Scodeggio, M., Gavazzi, G., Belsole, E., Pierini, D., \& Boseilli,
A. 1998, MNRAS, 301, 1001

\reference{color gradients}
Scodeggio, M. 2001, AJ, 121, 2413

\reference{}
Stanford, S.A., Eisenhardt, P.R., \& Dickinson, M. 1998, ApJ, 492, 461

\reference{}
Tonry, J.L. 1998, AJ, 115, 1

\reference{}
Tonry, J.L., \& Kochanek, C.S. 1999, AJ, 117, 2034

\reference{}
Tonry, J.L., \& Kochanek, C.S. 2000, AJ, 119, 1078


\reference{}
Trager, S.C., Faber, S.M., Worthey, G., \& Gonz\'alez, J.J. 2000, AJ,
119, 1645


\reference{Treu 2001} 
Treu, T., Stiavelli, M., Bertin, G., Casertano, S., \& Moller,
P. 2001, MNRAS, 326, 237

\reference{Treu 2002} 
Treu. T., Stiavelli, M., Casertano, S., Moller, P., \&
Bertin, G. 2002, ApJL, 564, L13

\reference{2016 KT}
Treu, T., \& Koopmans, L.V.E. 2002a, ApJ, 575, 87

\reference{1115}
Treu, T., \& Koopmans, L.V.E. 2002b, MNRAS, 337L, 6

\reference{} 
Turner, E.L., Ostriker, J.P., \& Gott, J.R., III 1984, ApJ, 284, 1


\reference{vDF}
van Dokkum, P.G., \& Franx, M. 1996, MNRAS, 281, 985

\reference{ cluster galaxies}
van Dokkum, P.G., Franx, M., Kelson, D.D., \& Illingworth, G.D. 1998,
ApJL, 504, L17

\reference{cluster interactions} 
van Dokkum, P.G., Franx, M., Fabricant, D., Kelson, D.D., \&
Illingworth, G.D. 1999, ApJL, 520, L95

\reference{progenitor bias} 
van Dokkum, P.G., Franx, M., Fabricant, D., Illingworth, G.D., \&
Kelson, D.D. 2000, ApJ, 541, 95

\reference{vDf2001 morphological verse star age}
van Dokkum, P.G., \& Franx, M. 2001, ApJ, 553, 90

\reference{field galaxies} 
van Dokkum, P.G., Franx, M., Kelson, D.D., \& Illingworth, G.D. 2001,
ApJL, 553, L39

\reference{}
Vogt, N.P., Forbes, D.A., Phillips, A.C., Gronwall, C., Faber, S.M.,
Illingworth, G.D., \& Koo, D.C. 1996, ApJL, 465, L15

\reference{0957}
Walsh, D., Carswell, R.F., \& Weymann, R.J. 1979, Nature, 279, 381

\reference{0047 discovery} 
Warren, S.J., Hewett, P.C., Lewis, G.F., Moller, P., Iovino, A., \&
Shaver, P.A. 1996, MNRAS, 278, 139

\reference{Clustering 1}
White, S.D.M., \& Rees, M.J. 1978, MNRAS, 183, 341

\reference{}
Worthey, G. 1994, ApJS, 95, 107

\reference{}
Wyithe, J.S.B., \& Loeb, A. 2002, ApJ, 577 57

\reference{}
Ziegler, B.L., et al.\ 2002, ApJL, 564, L69



\end{references}
\end{document}